\documentclass{article}
\usepackage[utf8]{inputenc}

\usepackage{graphicx}
\usepackage{amsmath}
\usepackage{amsfonts}
\usepackage{amssymb}
\usepackage{bm}
\usepackage{graphicx}
\usepackage{xcolor}
\usepackage{subcaption}
\usepackage{booktabs}
\usepackage[ruled,vlined]{algorithm2e}
\usepackage{subcaption}
\usepackage{natbib}

\SetKwInput{KwInput}{Input}                
\SetKwInput{KwOutput}{Output}              

\newcommand{\bmu}{\boldsymbol{\mu}}

\newcommand{\bSigma}{\boldsymbol{\Sigma}}

\DeclareMathOperator*{\argmax}{arg\,max}

\title{Robust variable selection for model-based learning in presence of adulteration}

\author{Andrea Cappozzo \footnote{Department of Statistics and Quantitative Methods, University of Milano-Bicocca, \texttt{andrea.cappozzo@unimib.it, francesca.greselin@unimib.it}}         \and
        Francesca Greselin\footnotemark[\value{footnote}] \and
        Thomas Brendan Murphy \footnote{School of Mathematics \& Statistics and Insight Research Centre, University College Dublin, \texttt{brendan.murphy@ucd.ie}}
}

\begin{document}
\date{\vspace{-5ex}}
\maketitle

\begin{abstract}
The problem of identifying the most discriminating features when performing supervised learning has been extensively investigated. In particular, several methods for variable selection in model-based classification have been proposed. Surprisingly, the impact of outliers and wrongly labeled units on the determination of relevant predictors has received far less attention, with almost no dedicated methodologies available in the literature. In the present paper, we introduce two robust variable selection approaches: one that embeds a robust classifier within a greedy-forward selection procedure and the other based on the theory of maximum likelihood estimation and irrelevance. The former recasts the feature identification as a model selection problem, while the latter regards the relevant subset as a model parameter to be estimated. The benefits of the proposed methods, in contrast with non-robust solutions, are assessed via an experiment on synthetic data. An application to a high-dimensional classification problem of contaminated spectroscopic data concludes the paper. 
\end{abstract}
\noindent
\section{Introduction} \label{intro}
Nowadays, in many scientific domains such as chemometrics, computer vision, engineering and genetics among others, it is increasingly common to measure hundreds or thousands of variables on each sample. In principle, depending on the problem at hand, all the available features might be relevant and thus deemed to be included in a subsequent analysis. Most often, however, incorporating every piece of information at our disposal unnecessarily increases model complexity and, ultimately, it may undermine the entire output of a statistical procedure. Model-based methods are particularly sensitive to the well-known \textit{curse of dimensionality} \citep{bellman1957dynamic}, as such models are over-parametrized and suffer from identifiability problems in high dimensional spaces \citep[][Chapter 8]{Bouveyron2014a,  bouveyron2019model}. Therefore, in a discriminant analysis context, selecting the useful variables that better unveil the group structure is crucial to learn an efficient classifier. This has been known for a long time, as demonstrated by the specific literature reviews on the topic in the fields of machine learning \citep{Blum1997, Yu2004, liu2007computational}, data mining \citep{Dash1997, Kohavi1997}, bioinformatics \citep{Saeys2007}, genomic \citep{yu2008feature} and statistics \citep{mclachlan2004discriminant, guyon2007causal, Fop2017b}. Nonetheless, the impact that outliers and wrongly labeled units cause on the efficient determination of discriminant variables has received far less attention. Indeed, the presence of attribute and class noise can heavily damage a classifier performance \citep{Zhu}, and most variable selection methods rely on the implicit assumption of dealing with an uncontaminated training set. 

In order to overcome this limitation, the present paper proposes two approaches for robust variable selection in model-based classification: one that embeds a robust classifier, recently introduced in the literature, in a greedy-forward stepwise procedure for model selection (Section \ref{sec:SRUW}); and the other based on the theory of maximum likelihood and the notion of irrelevant variables within robust ML estimation of normal mixtures (Section \ref{sec:EMST}). Both procedures rely on impartial trimming \citep{Gordaliza1991}: an appealing technique for robust parameter estimation in which no model assumption is a-priori required for the noise component. By leaving the anomalous units unmodeled, great flexibility is achieved and thus very heterogeneous contamination patterns can be effectively dealt with.

The remaining of the article is structured as follows. Section \ref{sec:feature_selection} formally characterizes the problem of variable selection in model-based discriminant analysis. 
In Section \ref{sec:REDDA_varsel}, the main features of the Robust Eigenvalue Decomposition Discriminant Analysis (REDDA) are reviewed. Two novel variable selection techniques resistant to outliers and label noise are introduced in  Section \ref{sec:robust_var_sel}: they are the main contributions of the present manuscript. Section \ref{sec:sim_study_varsel} is devoted to the comparison of several feature selection procedures within two simulation studies in an artificially contaminated scenario. Section \ref{sec:application_varsel} presents a high-dimensional discrimination study where our proposals for robust variable selection are successfully applied to a chemometrics contest. Section \ref{sec:conclusion_varsel} concludes the paper outlying some remarks and future research directions. Technical issues and computational details for the two novel methods are respectively deferred to \ref{sec:appendix_A}  and \ref{sec:appendix_B}.

\section{The problem of feature selection in discriminant analysis} \label{sec:feature_selection}
\begin{figure}
\includegraphics[scale=.7]{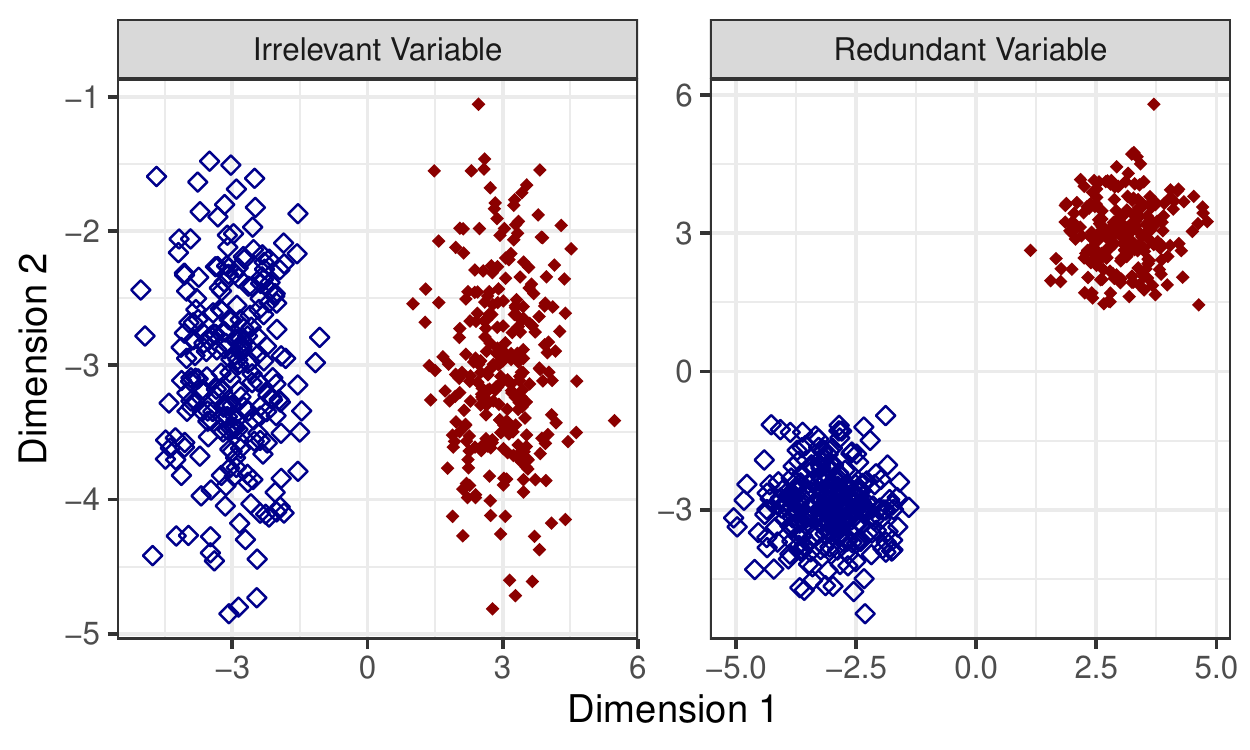}
\centering
\caption{Examples of learning scenarios for which the second dimension is irrelevant (left panel) or redundant (right panel) in discriminating the two groups.}
\label{fig:ex_rel_irrel}
 \end{figure}
The detection of $p$ relevant features (out of the whole collection of $P\gg p$ available variables) on which to train the classifier is particularly desirable, as \citep{mclachlan2004discriminant}:
\begin{itemize}
\item it simplifies parameter estimation and interpretation;
\item it avoids loss on predictive power due to the inclusion of irrelevant and redundant information;
\item it leads to cost reduction on future data collection and processing.
\end{itemize}
Therefore, with the aim of choosing the best predictors, it is crucial to define the concept of ``relevant variable''. The framework of model-based discriminant analysis allows to define ``relevance'' in terms of probabilistic dependence (or independence) with respect to the class membership \citep{Ritter2015}. The distribution of the \textit{relevant variables}, i.e., features that bring significant information on class separation, directly depends on the class membership itself. In discriminating men and women of the same ethnicity for example, the height is naturally relevant. \textit{Irrelevant or noisy variables}, on the contrary, do not contain any discriminating power, and hence their distribution is completely independent from the group structure. To continue with our previous example, hair and eye color do not convey any information on the gender of a person. Lastly, \textit{redundant variables} essentially contain discriminant information that is already provided by the relevant ones: their distribution is conditionally independent of the grouping variable, given the relevant ones. If the height of a person is known, little extra information is gained by finding out his/her head circumference for determining his/her gender. In Figure \ref{fig:ex_rel_irrel}, the first dimension is a relevant variable for discriminating the two groups, while the second dimension is respectively irrelevant in the left panel and redundant in the right one.

Depending on how the variable selection process interacts with the model estimation, two general approaches for feature identification can be defined. Following the nomenclature introduced by \cite{John1994}, \textit{filter methods} are those in which the selection acts as a pre (or post) processing step, discarding variables whose distribution appears non-informative. Since the selection via filter methods is performed separately from the model estimation, i.e., without reference to the class membership, such techniques may miss important grouping information; a standard example being Principal Component Analysis \citep{chang1983using}. For a state-of-the-art benchmark study on the comparison of filter methods for feature selection in  high dimensional classification, the reader is referred to \cite{Bommert2019}.

For the second class of methods the feature identification is ``wrapped'' around the classification procedure; hence they are denoted as \textit{wrapper approaches}. Within this framework, variable selection and model estimation are simultaneously performed, aiming at identifying the predictors that better describe the underlying data partition. Focusing on the model-based methods for classification, \cite{Murphy2012} provide a wrapper approach for feature selection in semi-supervised discriminant analysis, recasting the feature identification as a model selection problem. The authors develop a greedy search and a headlong search algorithm for finding a local optimum in the model space, inspired by the seminal work on variable selection in model-based clustering of \cite{Dean2006}, wherein for the first time the potential correlation between relevant and irrelevant variables is taken into account. Similarly, a general methodology for selecting predictors in model-based discriminant analysis is introduced in \cite{Maugis2011}, where also theoretical results on model identifiability and consistency of the proposed criterion are validated. More recently, a regularization approach for feature selection in model-based clustering and classification is introduced in \cite{Celeux2019}, where a lasso-like procedure is employed for overcoming the slowness yielded by stepwise algorithms when dealing with high-dimensional problems. The \texttt{SelvarMix R} package provides an efficient \texttt{C++} implementation of the afore-mentioned procedure. Unfortunately, no one of the wrapper methods listed here provide protection against outliers and label noise: the presence of only few adulterated data points can severely undermine the variable selection results (see Section \ref{sec:sim_study_varsel}). 

Lastly, methods that lie in between the two approaches have also been developed in the literature. Such hybrid methods usually involve feature selection based on some measure of separability between groups, like the one introduced by \cite{Indahl2004}, specifically tailored for spectroscopic data, and the one proposed by \cite{Andrews2014}. Further, a series of techniques based on metaheuristic strategies for variable selection in discriminant analysis can be found in \cite{Pacheco2006}, while the method of \cite{Chiang2004} relies on a stochastic search based on genetic algorithms. In general, even though being more complex and computationally intensive, wrapper approaches provide better classification results and more accurate representation of the data generating process \citep{Kohavi1997}. For this reason, the present manuscript will focus on wrapper approaches: the novel methods introduced in Section \ref{sec:robust_var_sel} fall within this category.

An important consideration to be made regards existing approaches that already provide robust selection of variables. In linear discriminant analysis (LDA), early-stage wrapper methods consider the employment of stepwise procedures in testing for no additional information, like the stepwise MANOVA described in Section 12.3 of \cite{mclachlan2004discriminant}: these are usually based on the likelihood ratio test Wilks' $\Lambda$ statistic. By respectively employing M-estimates and MCD-estimates to obtain a robust version of the Wilks' $\Lambda$ statistics, \cite{Krusinska1988} and \cite{Todorov2007} develop LDA-based techniques for variable selection resistant to outliers. Nevertheless, to our best knowledge, wrapper methods that perform robust feature selection in a more general framework are still missing in the literature.

Prior to present our novel contributions for variable selection resistant to outliers and label noise, the Robust Eigenvalue Decomposition Discriminant Analysis (REDDA) model is briefly reviewed in the upcoming Section; for a thorough treatment the interested reader is referred to \cite{Cappozzo2019b}.

\section{Robust model-based discriminant analysis} \label{sec:REDDA_varsel}
Model-based discriminant analysis \citep{mclachlan2004discriminant, Fraley2002} is a probabilistic framework for supervised classification, in which a classifier is built from a complete set of $N$ learning observations (i.e., the training set):
\begin{equation}
(\mathbf{x}, \mathbf{l})=\left\{\left(\mathbf{x}_{1}, \mathbf{l}_{1}\right), \ldots,\left(\mathbf{x}_{N}, \mathbf{l}_{N}\right) ; \mathbf{x}_{n} \in \mathbb{R}^{P},\,\, \mathbf{l}_{n}=\{l_{n1},\ldots, l_{nG}\}' \in \{0,1\}^G; \: n=1,\ldots,N\right\}
\end{equation}
where $\mathbf{x}_{n}$ is a $P$-dimensional continuous predictor and $\mathbf{l}_{n}$ is its associated class label, such that $l_{ng}=1$ if observation $n$ belongs to group $g$ and $0$ otherwise, $g=1,\ldots, G$, with, $\sum_{g=1}^{G} l_{ng}=1 \: \forall n=1,\ldots, N$. Alternatively, for sake of brevity, we will also employ the notation $l_{n}=g$ to denote the class of the $n$-th observation. We assume that the prior probability of group $g$ is $\mathbb{P}(\mathbf{l}=g)=\tau_g$, with $\tau_g>0$ and $\sum_{g=1}^G\tau_g=1$. The $g$th class-conditional density is modeled with a $P$-dimensional Gaussian distribution with mean vector $\boldsymbol{\mu}_g \in \mathbb{R}^P$ and positive semi-definite covariance matrix $\boldsymbol{\Sigma}_g \in PD(P)$: $\mathbf{x}_n|\mathbf{l}_{n}=g \sim N_P(\boldsymbol{\mu}_g, \boldsymbol{\Sigma}_g)$. Therefore, the joint density of $(\mathbf{x}_n, \mathbf{l}_n)$ is given by:
\begin{equation} \label{joint_density_EDDA_varsel}
p(\mathbf{x}_n,\mathbf{l}_n; \boldsymbol{\theta}) = p(\mathbf{l}_n;\boldsymbol{\tau})p(\mathbf{x}_n|\mathbf{l}_n; \boldsymbol{\mu}_g, \boldsymbol{\Sigma}_g)=\prod_{g=1}^G \left[ \tau_g \phi(\mathbf{x}_n; \boldsymbol{\mu}_g, \boldsymbol{\Sigma}_g) \right]^{l_{ng}}
\end{equation}
where $\phi(\cdot; \boldsymbol{\mu}_g, \boldsymbol{\Sigma}_g)$ denotes the multivariate normal density and $\boldsymbol{\theta}$ is the collection of parameters to be estimated, $\boldsymbol{\theta}= \{ \tau_1, \ldots,  \tau_G, \boldsymbol{\mu}_1, \ldots, \boldsymbol{\mu}_G, \boldsymbol{\Sigma}_1, \ldots, \boldsymbol{\Sigma}_G \}$. Eigenvalue Decomposition Discriminant Analysis (EDDA) is a family of classifiers developed from the probabilistic structure in \eqref{joint_density_EDDA_varsel}, wherein different assumptions about the covariance matrices are considered. Particularly, EDDA is based on the following eigenvalue decomposition \citep{Banfield1993, Celeux1995}:
\begin{equation} \label{sigma_dec_varsel}
\boldsymbol{\Sigma}_g=\lambda_g\boldsymbol{D}_g\boldsymbol{A}_g\boldsymbol{D}^{'}_g
\end{equation}
where $\boldsymbol{D}_g$ is an orthogonal matrix of eigenvectors, $\boldsymbol{A}_g$ is a diagonal matrix such that $|\boldsymbol{A}_g|=1$ and $\lambda_g=|\boldsymbol{\Sigma}_g|^{1/p}$. These elements correspond respectively to the orientation, shape and volume (alternatively called scale) of the Gaussian components. Allowing each parameter in \eqref{sigma_dec_varsel} to be equal or different across groups, \cite{Bensmail1996} defined a family of 14 patterned models. 
\cite{Cappozzo2019b} introduced a robust modification to EDDA, hereafter denoted REDDA, in which parameter estimates are protected against label noise and outliers by means of a \textit{trimmed mixture log-likelihood} \citep{Neykov2007}:
\begin{equation} \label{trim_ll_REDDA_varsel}
\ell_{trim}(\boldsymbol{\tau}, \boldsymbol{\mu}, \boldsymbol{\Sigma}| \mathbf{X}, \mathbf{l})=
\sum_{n=1}^N \zeta(\mathbf{x}_n)\sum_{g=1}^G l_{ng} \log{\left(\tau_g \phi(\mathbf{x}_n; \boldsymbol{\mu}_g, \boldsymbol{\Sigma}_g)\right)}
\end{equation}
where \(\zeta(\cdot)\) is a 0-1 trimming indicator function, that expresses whether observation \(\mathbf{x}_n\) is trimmed off or not. A fixed fraction \(\gamma\) of observations is unassigned by setting \(\sum_{n=1}^N \zeta(\mathbf{x}_n)=\lceil N(1-\gamma)\rceil\). The \textit{labelled trimming level} \(\gamma\) accounts for possible adulteration, namely outliers and label noise, in the training set. Maximization of \eqref{trim_ll_REDDA_varsel} is carried out via a generalization of the FastMCD algorithm by \cite{Driessen1999}, adapted to deal with parsimonious structures in the covariance matrices. Particularly, in this context the \textit{Concentration step} (C-step) is enforced by temporarily discarding \(\lfloor N \gamma \rfloor\) units with lowest value of:
\begin{equation}\label{cond_dens_var_sel}
\phi\left(\mathbf{x}_n; \hat{\boldsymbol{\mu}}_g, \hat{\boldsymbol{\Sigma}}_g \right) \:\:\:\:\: g=1,\ldots, G.
\end{equation}
For these observations, $\zeta(\mathbf{x}_n)=0$ in \eqref{trim_ll_REDDA_varsel} as they will not be accounted for in the  next estimation step: the algorithm stops once the less plausible $\lfloor N \gamma \rfloor$ discarded units, out of the $N$ units in the learning set, are confirmed to be the same on two consecutive iterations. Notice that the mixing proportions $\hat{\tau}_g$ do not appear in \eqref{cond_dens_var_sel}: the estimated group-conditional densities act as discriminative tools for trimming, so that \(\lfloor N \gamma \rfloor\) overall samples are removed at each step. At the end of the procedure, a value of $\zeta(\mathbf{x}_n)=0$ corresponds to identify $\mathbf{x}_n$ as an unreliable unit. The REDDA classifier can then be employed for assigning an unlabeled sample $\mathbf{y}_m$, $m=1,\ldots, M$ (i.e., the test set) to the class $g$ whose associated posterior probability
\begin{equation}\label{MAP_varsel}
\hat{z}_{mg}=\frac{\hat{\tau}_g \phi(\mathbf{y}_m; \hat{\boldsymbol{\mu}}_g, \hat{\boldsymbol{\Sigma}}_g)}{\sum_{j=1}^G\hat{\tau}_j \phi(\mathbf{y}_m; \hat{\boldsymbol{\mu}}_j, \hat{\boldsymbol{\Sigma}}_j)}.
\end{equation}
is highest, by means of the usual maximum a posteriori (MAP) rule. In addition, also the trimmed units can be a-posteriori assigned to the component \(g\) displaying the highest value of $\hat{\tau}_g \phi(\mathbf{x}_n;\hat{\boldsymbol{\mu}}_g,\hat{\boldsymbol{\Sigma}}_g)$, to recover a reasonable label for observations that previously got an adulterated one.

\section{Robust Variable Selection in model-based classification} \label{sec:robust_var_sel}
In the present Section we introduce two novel wrapper approaches for robust variable selection in high-dimensional model-based classification.

In Section \ref{sec:SRUW}, the REDDA method is embedded in a greedy-forward procedure for model selection. A robust classification rule is constructed in a step-wise manner, by considering the inclusion of extra variables and also the removal of existing variables to/from the model, conditioning on their discriminating power. Particularly, the selection procedure is based on a robust information criterion, that accounts for the possible presence of outliers and label noise in the dataset.

In Section \ref{sec:EMST}, the theory of maximum likelihood estimation and the notion of irrelevant variables for normal mixtures is employed for defining a ML subset selector, along the lines of the procedure introduced in section 5.3.3 of \cite{Ritter2015} for the unsupervised framework. The identification of the relevant subset is regarded as a parameter to be estimated via ML: an algorithmic procedure is derived for maximizing the objective function. The Section concludes with a comparison, highlighting strengths and weaknesses of the two proposals.
\subsection{The robust stepwise greedy-forward approach via TBIC} \label{sec:SRUW}
\begin{figure}[h]
\centering
\includegraphics[scale=0.8]{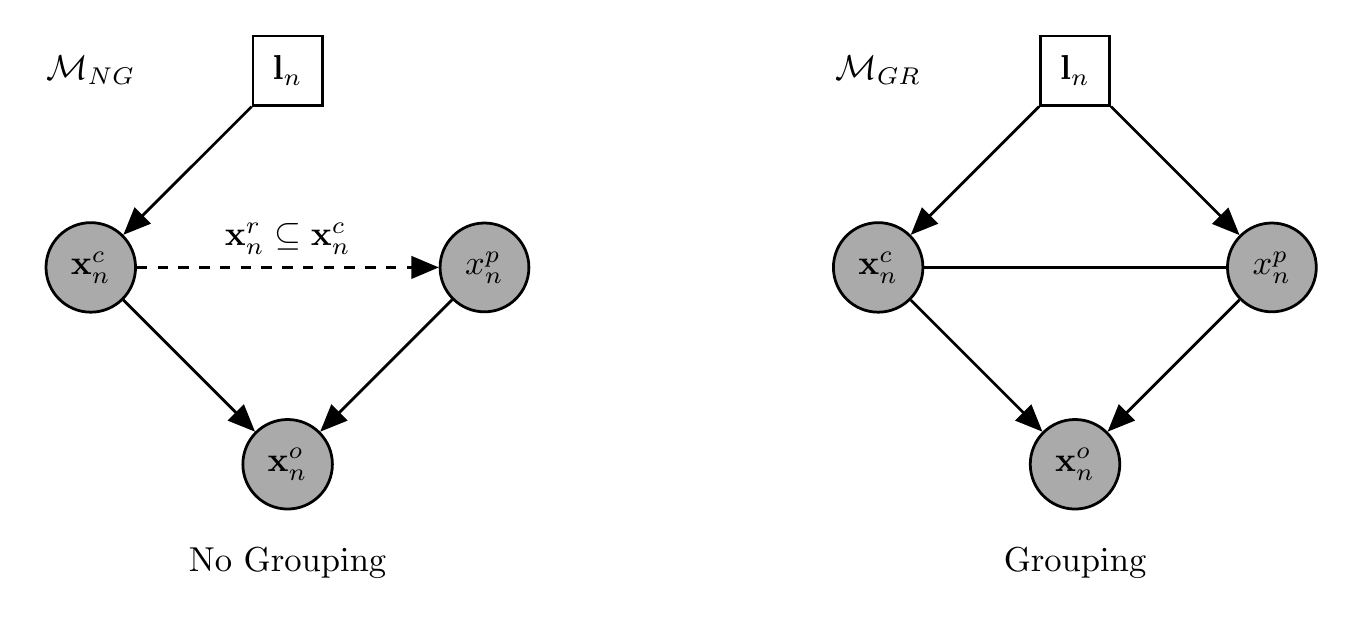}
\caption{Graphical Representation of the Grouping and the No Grouping models}
\label{fig:var_sel}
\end{figure}

The present procedure searches for the set of relevant variables in a greedy-stepwise manner. That is, we start from the empty set and we sequentially add relevant variables until no more discriminating features are available.
More specifically, following the notation introduced in Section \ref{sec:REDDA_varsel}, in each step of the algorithm we partition the learning observations $\mathbf{x}_n$, $n=1,\ldots,N$, into three parts $\mathbf{x}_n=(\mathbf{x}_n^{c},x_n^{p},\mathbf{x}_n^{o})$, where:
\begin{itemize}
\item $\mathbf{x}_n^{c}$ indicates the set of variables currently included in the model,
\item $x_n^{p}$ the variable proposed for inclusion,
\item $\mathbf{x}_n^{o}$ the remaining variables.
\end{itemize}
In order to decide whether to include the proposed variable $x_n^{p}$, we compare the following two competing models:
\begin{itemize}
\item \textit{Grouping ($\mathcal{M}_{GR}$):} 
\[p(\mathbf{x}_n|\mathbf{l}_n)=p(\mathbf{x}_n^{c},x_n^{p},\mathbf{x}_n^{o}|\mathbf{l}_n)=
p(\mathbf{x}_n^{c},x_n^{p}|\mathbf{l}_n)p(\mathbf{x}_n^{o}|x_n^{p},\mathbf{x}_n^{c})\]
\item \textit{No Grouping ($\mathcal{M}_{NG}$):} 
\[p(\mathbf{x}_n|\mathbf{l}_n)=p(\mathbf{x}_n^{c},x_n^{p},\mathbf{x}_n^{o}|\mathbf{l}_n)=p(\mathbf{x}_n^{c}|\mathbf{l}_n)
p(x_n^{p}|\mathbf{x}_n^{r} \subseteq \mathbf{x}_n^{c})p(\mathbf{x}_n^{o}|x_n^{p},\mathbf{x}_n^{c})\]
\end{itemize}
where $\mathbf{x}_n^{r}$ denotes a subset of the currently included variables $\mathbf{x}_n^{c}$. The grouping model specifies that $x_n^{p}$ provides extra grouping information beyond that provided by $\mathbf{x}_n^{c}$; whereas the No Grouping model specifies that $x_n^{p}$ is conditionally independent of the group membership given $\mathbf{x}_n^{r}$. The reason for considering $\mathbf{x}_n^{r}$ in the conditional distribution being that $x_n^{p}$ might be related to only a subset of the grouping variables $\mathbf{x}_n^{c}$ \citep{Maugis2009, Maugis2009a, Maugis2011}.   The differences between the two models are graphically illustrated in Figure \ref{fig:var_sel}.
The model structure of $p(\mathbf{x}_n^{o}|x_n^{p},\mathbf{x}_n^{c})$ is assumed to be the same for both grouping and no grouping specification, and we let  $p(\mathbf{x}_n^{c},x_n^{p}|\mathbf{l}_n)$ and $p(\mathbf{x}_n^{c}|\mathbf{l}_n)$ be a normal density with parsimonious covariance structure, according to the model assumptions introduced in the previous Section. 
Additionally, we assume $p(x_n^{p}|\mathbf{x}_n^{r} \subseteq \mathbf{x}_n^{c})$ to be a normal linear regression model, as a result from conditional multivariate normal means. The selection of which model to prefer is carried out employing a robust approximation to the Bayes Factor. More specifically, the Bayes Factor \citep{Kass1995} is equal to the ratio between the integrated likelihood of the two competing models:
\begin{equation} \label{BF}
\mathcal{B}_{GR,NG}=\frac{p(\mathbf{x}_n|\mathcal{M}_{GR})}{p(\mathbf{x}_n|\mathcal{M}_{NG})}=\frac{\int p(\mathbf{x}_n|\boldsymbol{\theta}_{GR},\mathcal{M}_{GR})p(\boldsymbol{\theta}_{GR}|\mathcal{M}_{GR})d \boldsymbol{\theta}_{GR}}{\int p(\mathbf{x}_n|\boldsymbol{\theta}_{NG},\mathcal{M}_{NG})p(\boldsymbol{\theta}_{NG}|\mathcal{M}_{NG})d \boldsymbol{\theta}_{NG}}
\end{equation}  
where $\boldsymbol{\theta}_{GR}$ and $\boldsymbol{\theta}_{NG}$ denote the set of parameters for the Grouping ($GR$) and the No Grouping ($NG$)  model, respectively. When no prior preference for one of the two models is considered, \eqref{BF} is equal to the posterior odds in favour of $\mathcal{M}_{GR}$. The Bayes Factor can therefore be used for assessing to which extent the data supports the $GR$ structure compared to the $NG$ formulation. Along the lines of \cite{Raftery}, the Bayesian Information Criterion
\[BIC=2\times log\:\:maximized\:\:likelihood-v \log{N}\]
is used as an approximation for the integrated likelihood, where $v$ is a penalty term (number of parameters in the model) and $N$ is the sample size \citep{Schwarz1978}. Thus, twice the logarithm of $\mathcal{B}_{GR,NG}$ can be approximated with
\begin{equation} \label{bic_diff}
2\log{\left(\mathcal{B}_{GR, NG}\right)}\approx BIC(GR)-BIC(NG)
\end{equation}
and a variable $x_n^{p}$ with a positive difference in $BIC(GR)-BIC(NG)$
is a candidate for being added to the model. For avoiding the detrimental effect that class and attribute noise might produce in the variable selection procedure, the Trimmed BIC (TBIC), firstly introduced in \cite{Neykov2007}, is employed as a robust proxy for the quantities in \eqref{bic_diff}. Let us define: 
\begin{align} \label{TBIC_gr}
\begin{split}
TBIC(GR) &= \underbrace{2\sum_{n=1}^N \zeta(\mathbf{x}_n^{c}, x_n^p)\sum_{g=1}^G l_{ng} \log{\left(\hat{\tau}_g^{cp} \phi(\mathbf{x}_n^{c}, x_n^p; \hat{\boldsymbol{\mu}}_g^{cp}, \hat{\boldsymbol{\Sigma}}_g^{cp})\right)}}_{2 \times \text{trimmed log maximized likelihood of }p(\mathbf{x}_n^{c},x_n^{p},\mathbf{l}_n)}
- v^{cp} log(N^*)
\end{split}
\end{align}
%
\begin{align} \label{TBIC_no}
\begin{split}
TBIC(NG) &= \underbrace{2\sum_{n=1}^N \iota(\mathbf{x}_n^{c}, x_n^p)\sum_{g=1}^G l_{ng} \log{\left(\hat{\tau}_g^{c} \phi(\mathbf{x}^c_n; \hat{\boldsymbol{\mu}}_g^{c}, \hat{\boldsymbol{\Sigma}}_g^{c})\right)}}_{2 \times \text{trimmed log maximized likelihood of }p(\mathbf{x}_n^{c}, \mathbf{l}_n)} - v^{c} log(N^*)+\\
&\underbrace{+2\sum_{n=1}^N \iota(\mathbf{x}_n^{c}, x_n^p) \log{\left[\phi\left(x_n^{p}; \hat{\alpha} + \hat{\boldsymbol{\beta}}^{'}\mathbf{x}_n^{r},\hat{\sigma}^2\right)\right]}}_{2 \times \text{trimmed log maximized likelihood of }p(x_n^{p}|\mathbf{x}_n^{r} \subseteq \mathbf{x}_n^{c})} - v^{p}log(N^*).
\end{split}
\end{align}
The penalty terms $v^{cp}$ and $v^{c}$ indicate the number of parameters for a REDDA model respectively estimated on the set of variables $\mathbf{x}_n^c, x_n^p$ and $\mathbf{x}_n^c $; while $v^{p}$ accounts for the number of parameters in the linear regression of $x_n^{p}$ on $\mathbf{x}_n^{r}$. The 0-1 indicator functions $\zeta(\cdot)$ and $\iota(\cdot)$ identify the subset of observations that have null weight in the trimmed likelihood under the grouping and no grouping models, with $N^*=\sum_{n=1}^N \zeta(\mathbf{x}_n)=\sum_{n=1}^N \iota(\mathbf{x}_n)$.


In detail, the parameters $\left\{ \tau_g^{cp}, \: \boldsymbol{\mu}_g^{cp} \: \boldsymbol{\Sigma}_g^{cp} \right\},\:g=1,\ldots,G$ of the grouping model are estimated through a standard REDDA fitted on the variables $\mathbf{x}_n^c, x_n^p$, in which the C-step is enforced discarding the \(\lfloor N \gamma \rfloor\) samples with lowest value of
\begin{equation} \label{D-group}
D_{Grouping}\left(\mathbf{x}_n^{c}, x_n^p; \hat{\boldsymbol{\theta}}_{GR} \right)=\sum_{g=1}^Gl_{ng}\log{ \left[ \phi \left(\mathbf{x}_n^{c}, x_n^p; \hat{\boldsymbol{\mu}}^{cp}_g, \hat{\boldsymbol{\Sigma}}^{cp}_g \right)\right]} \:\:\:\:\: n=1,\ldots,N
\end{equation}
likewise for the general case in \eqref{cond_dens_var_sel}. For the no grouping model, REDDA needs to be fitted only on the set of currently included variables $\mathbf{x}_n^{c}$, coupled with the linear regression of $x_n^{p}$ on $\mathbf{x}_n^{r}$. For this case, the discriminating function reads:
\begin{align}  \label{D_NOgroup_2}
\begin{split}
D_{No\:\:Grouping}\left(\mathbf{x}_n^{c}, x_n^p; \hat{\boldsymbol{\theta}}_{NG} \right)&=\sum_{g=1}^G l_{ng}\log{ \left[ \phi \left(\mathbf{x}^c_n; \hat{\boldsymbol{\mu}}^{c}_g, \hat{\boldsymbol{\Sigma}}^{c}_g \right)\right]} +\log{\left[ \phi(x^p_n; \hat{\alpha}+\hat{\boldsymbol{\beta}}^{'}\mathbf{x}^r_n, \hat{\sigma}^2)\right]}
\end{split}
\end{align}
for $n=1,\ldots,N$. That is, at each iteration of the procedure that leads to the final robust estimates, we discard \(\lfloor N\gamma \rfloor\) samples with the lowest contribution to the conditional likelihood under the no grouping model. Once the C-step is enforced, the set of parameters $\left\{\alpha, \: \boldsymbol{\beta}, \: \sigma^2  \right\}$ for the regression part is robustly estimated via ML on the untrimmed observations, in which a stepwise method is employed for automatically choosing the subset of regressors $\mathbf{x}^r_n$. Further details concerning the implementation are included in \ref{sec:appendix_A}. 

After each addition stage, we make use of the same procedure described above to check whether an already chosen variable in $\mathbf{x}_n^c$ should be removed: in this case $x_n^p$ takes the role of the variable to be dropped, and a negative difference in terms of TBIC implies the exclusion of $x_n^p$ to the set of currently included variables. The procedure iterates between variable addition and removal stage until two consecutive steps have been rejected, then it stops. Notice that, whenever $\gamma=0$, BIC and TBIC coincide and the entire approach reduces to the methodology described in \cite{Maugis2011}.

A last worthy note regards the theoretical justification for the employment of TBIC as an approximation of the integrated likelihood. The rationale arises from the spurious outliers model, firstly introduced in \cite{Gallegos2005}, as the probabilistic specification for the contaminated sub-sample. Let $q_n$ denote an indicator of genuine observations, such that $q_n=1$ when $\{(\mathbf{x}_n, \mathbf{l}_n) \}$ is a ``regular"  unit and $q_n=0$ whenever $\{(\mathbf{x}_n, \mathbf{l}_n) \}$ presents some sort of contamination/adulteration. Notice that the complete observation $\{(\mathbf{x}_n, \mathbf{l}_n) \}$ might be regarded as an outlier whenever either the associated label and/or some of its predictors present unusual values. In such a way, we account for both attribute and class noise.
The data generating distribution for a specific observation $\{(\mathbf{x}_n, \mathbf{l}_n) \}$ is then assumed to be as follows:
\begin{equation} \label{sp_out}
p(\mathbf{x}_n,\mathbf{l}_n|q_n;\boldsymbol{\theta})= p(\mathbf{x}_n, \mathbf{l}_n; \boldsymbol{\theta})^{q_n } w(\mathbf{x}_n, \mathbf{l}_n; \boldsymbol{\psi}_n)^{(1-q_n)}
\end{equation}
where $p(\mathbf{x}_n, \mathbf{l}_n; \boldsymbol{\theta})$ denotes the probability distribution for the regular bulk of the data, in our context being alternatively the Grouping or the No Grouping model; and $w(\mathbf{x}_n, \mathbf{l}_n;\boldsymbol{\psi}_n)$ is an almost arbitrary, subject specific probability density function, parametrized by $\boldsymbol{\psi}_n \in \boldsymbol{\Psi}_n$. For an independent sample of $N$ observations, the likelihood for the model in \eqref{sp_out} is therefore given by:
\begin{equation} \label{sp_out_ll}
\prod_{n=1}^N p(\mathbf{x}_n, \mathbf{l}_n; \boldsymbol{\theta})^{q_n } \prod_{n=1}^N w(\mathbf{x}_n, \mathbf{l}_n;\boldsymbol{\psi}_n)^{(1-q_n)}
\end{equation}
where a fixed $\gamma \%$ of contamination is assumed such that $N^*=\sum_{n=1}^N q_n=\lceil N(1-\gamma)\rceil$. 
 
Let $\mathcal{N}=\{N_1,N_0\}$ be a partition of $N$ into regular and non-regular observations, indexed by $q_n$ being either $1$ or $0$ for $n=1,\ldots,N$, with $|N_1|=\lceil N(1-\gamma)\rceil$ and $|N_0|=\lfloor N\gamma \rfloor$, respectively. Further, denote with $\mathcal{D}(N)$ the set of all partitions of such type, with $|\mathcal{D}(N)|={N \choose \lceil N(1-\gamma)\rceil}$. The non-regular contribution of the contaminated observations can be avoided in maximizing \eqref{sp_out_ll} with respect to $\boldsymbol{\theta}$ when the $w(\cdot;\boldsymbol{\psi}_n)$s satisfy
\begin{equation} \label{sep_condition}
\argmax_{\mathcal{N}\in \mathcal{D}(N)} \max_{\boldsymbol{\theta}}\prod_{n=1}^N p(\mathbf{x}_n, \mathbf{l}_n; \boldsymbol{\theta})^{q_n } \subseteq \argmax_{\mathcal{N}\in \mathcal{D}(N)} \max_{\boldsymbol{\psi}_1,\ldots,\boldsymbol{\psi}_N}\prod_{n=1}^N w(\mathbf{x}_n, \mathbf{l}_n;\boldsymbol{\psi}_n)^{(1-q_n)}.
\end{equation}
The condition in \eqref{sep_condition} means that the configuration that maximizes the first term in \eqref{sp_out_ll} automatically maximizes the second one \citep{Gallegos2005}. More specifically, the partitions assigning $\lceil N(1-\gamma)\rceil$ regular units 
that maximize the likelihood of the genuine observations are contained in the set of partitions assigning $\lfloor N\gamma \rfloor$ non regular
units 
that maximize the likelihood corresponding to the noise.
Condition \eqref{sep_condition} holds under general and non-restrictive assumptions on the non regular units, particularly, $w(\cdot;\boldsymbol{\psi}_n)$ can easily accommodate observations that can be merely regarded as outliers \citep{Gallegos2005, Garcia-Escudero2008}. The contaminated observations are therefore no more considered in the estimation process, and the model log-likelihood simplifies to:
\begin{equation} \label{obj_trim_fun}
\sum_{n=1}^N q_n\log{p(\mathbf{x}_n, \mathbf{l}_n; \boldsymbol{\theta})}
\end{equation}
to be maximized with respect to the set of parameters $\boldsymbol{\theta}$; details are reported in \ref{sec:appendix_A}. Finally, the integrated log-likelihood for \eqref{obj_trim_fun} can be approximated via the Bayesian Information Criterion:
\begin{equation} \label{bic_trim}
2\sum_{n=1}^Nq_n\log{p(\mathbf{x}_n, \mathbf{l}_n; \hat{\boldsymbol{\theta}})}-v\log{N^*}
\end{equation} 
where $\hat{\boldsymbol{\theta}}$ denotes MLE for the simplified log-likelihood, $v$ is the number of parameters and $N^*$ is the number of data values that contribute to the summation in \eqref{obj_trim_fun} \citep{Kass1993}. Depending which scenario is considered, \eqref{bic_trim} defines \eqref{TBIC_gr} or \eqref{TBIC_no} under the Grouping and the No Grouping model, respectively.

\subsection{The ML subset selector approach} \label{sec:EMST}
The second approach we consider for robust variable selection in model-based classification stems from the maximum likelihood subset selector theory developed for clustering, where the main reference is Section 5.3.3 of \cite{Ritter2015}. Particularly, being classification a generally simpler problem than unsupervised learning, the ML subset selection ideas are naturally adapted to a robust supervised context with variable selection. Here we build a model for the entire $P$-dimensional space in which the observations lie, exploiting theoretical results for the conditional distribution of the multivariate Gaussian under irrelevance. Let us introduce the following notation: for $\boldsymbol{\Sigma} \in PD(P)$, denote its restriction to the variables in $F\subseteq 1,\ldots,P$ by $\boldsymbol{\Sigma}_F$, with size $|F|=p$. The block-wise representation of $\bSigma$, via the natural order of $F$, is therefore:
\[ \bSigma = \begin{pmatrix}
  \bSigma_F & \bSigma_{F,E}\\
  \bSigma_{E,F} & \bSigma_E
  \end{pmatrix} \]
with $E=\bar{F}$ and $|E|=P-p$. Analogously, the vector $\boldsymbol{\mu}_F$ is the projection of $\boldsymbol{\mu}\in \mathbb{R}^P$ onto the variables in $F$, following the natural order of $F$. 
For a generic observation $\mathbf{x}_n \in \mathbb{R}^P$, the canonical projection of a normal distribution to a subset $F$ of variables is described by the restrictions $\boldsymbol{\mu}_F$ and $\boldsymbol{\Sigma}_F$ of its parameters, with the equality $N_{\boldsymbol{\mu}, \boldsymbol{\Sigma}}(\mathbf{x}_{n,F})=N_{\boldsymbol{\mu}_{F}, \boldsymbol{\Sigma}_F}(\mathbf{x}_{n,F})$ such that $\mathbf{x}_{n,F}\sim N(\boldsymbol{\mu}_{F}, \boldsymbol{\Sigma}_F)$. Considering the notation introduced in Section \ref{sec:REDDA_varsel} and applying standard results for multivariate normal theory, (see, for example, Theorem 3.2.4 in \cite{bibby1979multivariate}), the conditional distribution of $\mathbf{x}_{n,E}$ given $\mathbf{x}_{n,F},\mathbf{l}_n$ reads:
\begin{equation}
\mathbf{x}_{n,E}|\mathbf{x}_{n,F},\mathbf{l}_n=g\sim \mathcal{N}_{P-p}\left( \bmu_{g,E|F}+\boldsymbol{G}_{g,E|F}\mathbf{x}_{n,F};\bSigma_{g,E|F}\right)
\end{equation}
where $\bmu_{g,E|F}=\bmu_{g,E}-\boldsymbol{G}_{g,E|F}\bmu_{g,F}$, $\bSigma_{g,E|F}=\bSigma_{g,E}-G_{g,E|F}\bSigma_{g,F,E}$ and $\boldsymbol{G}_{g,E|F}=\bSigma_{g,E,F}\bSigma_{g,F}^{-1}$, $g=1,\ldots,G$. Now assume that $E$ is an irrelevant subset with respect to $F$, that is, the class membership $\mathbf{l}_n$ is conditionally independent of $\mathbf{x}_{n,E}$ given $\mathbf{x}_{n,F}$. By Lemma 5.2 and Theorem 5.7 of \cite{Ritter2015}, the parameters $\boldsymbol{G}_{g,E|F}$, $\bmu_{g,E|F}$ and $\bSigma_{g,E|F}$ do not depend on class $g$; applying the product formula we thus obtain the following specification for the joint density of $(\mathbf{x}_{n,F}, \mathbf{x}_{n,E}, \mathbf{l}_n)$:
\begin{align} \label{prob_dens_EF}
\begin{split}
p(\mathbf{x}_{n,F}, \mathbf{x}_{n,E}, \mathbf{l}_n)&=p(\mathbf{x}_{n,F}, \mathbf{x}_{n,E}|\mathbf{l}_n)p(\mathbf{l}_n)=\\
&=p(\mathbf{x}_{n,E}|\mathbf{x}_{n,F}, \mathbf{l}_n)p(\mathbf{x}_{n,F}| \mathbf{l}_n)p(\mathbf{l}_n)=\\
&=p(\mathbf{x}_{n,E}|\mathbf{x}_{n,F})p(\mathbf{x}_{n,F}| \mathbf{l}_n)p(\mathbf{l}_n).
\end{split}
\end{align}
Therefore, for a sample of $N$ observations, drawn from the random variable $\mathbf{X}$, the associated trimmed log-likelihood for the probability density in \eqref{prob_dens_EF} is:

\begin{multline} \label{trim_ll_EMST_varsel}
\ell_{trim}(\boldsymbol{\tau}, \boldsymbol{\mu}_F, \boldsymbol{\Sigma}_F,\boldsymbol{G}_{E|F},\bmu_{E|F},\bSigma_{E|F}| \mathbf{X}, \mathbf{l})=\\
=\sum_{n=1}^N\zeta{(\mathbf{x}_n)} \left( \sum_{g=1}^G l_{ng} \log \left[ \tau_g \phi(\mathbf{x}_{n,F}; \boldsymbol{\mu}_{g,F}, \boldsymbol{\Sigma}_{g,F}) \right]+\right. \\
+\log \left[ \phi(\mathbf{x}_{n,E}-\boldsymbol{G}_{E|F}\mathbf{x}_{n,F};\bmu_{E|F},\bSigma_{E|F})\right] \bigg)
\end{multline}
where the identification of the relevant variables belonging to the subset $F$ is regarded as a model parameter. Maximization of \eqref{trim_ll_EMST_varsel} is carried out via a modification of the EMST algorithm introduced in \cite{Ritter2015}, adapted to the classification framework and extended to flexibly account for the entire family of patterned models of \cite{Bensmail1996}. The main steps involving the estimation procedure are given below, further details concerning the implementation can be found in \ref{sec:appendix_B}.
  \begin{enumerate}
\item
\emph{Robust Initialization:}
  \begin{itemize}
  \item If $N$ is sufficiently large compared to $P$ and $G$, draw a random $(P + 1)$-subset for each class $g$, $g=1,\ldots,G$. The first M-step will be computed only on such units: this is achieved by setting $\zeta(\mathbf{x}_n)=1$ if $\mathbf{x}_n$ belongs to the drawn subset, otherwise $\zeta(\mathbf{x}_n)=0$. Go to step 2 of the algorithm.
  \item If $N$ is small compared to $P$ and $G$, draw a random $(p + 1)$-subset for each class $g$, $g=1,\ldots,G$ and set $\zeta(\mathbf{x}_n)=1$ if $\mathbf{x}_n$ belongs to any of such $G$ subsets, otherwise set $\zeta(\mathbf{x}_n)=0$. 
  
  Draw a random subset $\hat{F}^{(0)}$ of dimension $p$ from $1,\ldots,P$ and compute:
\[  \hat{\boldsymbol{\mu}}_{g,\hat{F}^{(0)}}=\frac{\sum_{n=1}^N \zeta(\mathbf{x}_n)l_{ng}\mathbf{x}_{n,\hat{F}^{(0)}}}{\sum_{n=1}^N\zeta(\mathbf{x}_n)l_{ng}}\:\:\:\:\: g=1,\ldots, G,\]
and $\hat{\boldsymbol{\Sigma}}^{(0)}_{g,\hat{F}^{(0)}}$, $g=1,\ldots, G$, depending on the considered patterned model, refer to \cite{Bensmail1996} for the details. Lastly, update the trimming function $\zeta(\mathbf{x}_n)$, $n=1,\ldots,N$, setting $\zeta(\mathbf{x}_n)=0$ for the \(\lfloor N \gamma\rfloor \) samples with lowest value of
\[l_{ng} \log \left[\phi(\mathbf{x}_{n,F^{(0)}}; \hat{\boldsymbol{\mu}}^{(0)}_{g,F^{(0)}}, \hat{\boldsymbol{\Sigma}}^{(0)}_{g,F^{(0)}}) \right]\]
and $\zeta(\mathbf{x}_n)=1$ otherwise.
  \end{itemize}
\item (\textit{M-step})

Compute:
\[\hat{\tau}_g=\frac{\sum_{n=1}^N \zeta(\mathbf{x}_n)l_{ng}}{\lceil N(1-\gamma)\rceil}\:\:\:\:\: g=1,\ldots, G\]
\[  \hat{\boldsymbol{\mu}}_g=\frac{\sum_{n=1}^N \zeta(\mathbf{x}_n)l_{ng}\mathbf{x}_n}{\sum_{n=1}^N\zeta(\mathbf{x}_n)l_{ng}}\:\:\:\:\: g=1,\ldots, G.\]
Estimation of $\boldsymbol{\Sigma}_g$ depends on the considered patterned model, details are given in \cite{Bensmail1996}.

Notice that the estimates are computed for the full dimension $P$, that is $\hat{\bmu}_g \in \mathbb{R}^P$ and $\hat{\bSigma}_g \in PD(P)$, respectively. In addition, robustly compute also the pooled mean:
\[\hat{\bmu}=\frac{\sum_{n=1}^N \zeta(\mathbf{x}_n)\mathbf{x}_n}{\lceil N(1-\gamma)\rceil}.\]
Depending on the considered patterned model, formulae for the associated pooled estimate $\hat{\bSigma}$ are detailed in \ref{sec:appendix_B}. 

\item (\textit{S-step})

Minimize the difference:
\begin{equation} \label{s-step_varsel}
h(F)=\sum_{g=1}^G\hat{\tau}_g \log \det \hat{\bSigma}_{g,F}-\log \det \hat{\bSigma}_{F}
\end{equation}
w.r.t. the subset  $\hat{F}\subseteq 1,\ldots,P$, with $|\hat{F}|=p$, where $\hat{\bSigma}_{g,\hat{F}}$ is the restriction of $\hat{\bSigma}_{g}$ to $\hat{F}$. The minimization of \eqref{s-step_varsel} involves a discrete structure optimization, that becomes quickly unfeasible as ${P \choose p}$ grows: a genetic algorithm is proposed for solving it (more details in \ref{sec:appendix_B}).
\item (\textit{T-step})

Compute the MLE's for the regression parameters
\[\hat{\boldsymbol{G}}_{\hat{E}|\hat{F}}=\hat{\bSigma}_{\hat{E},\hat{F}}\hat{\bSigma}_{\hat{F}}^{-1}\]
\[\hat{\bmu}_{\hat{E}|\hat{F}}=\hat{\bmu}_{\hat{E}}-\hat{\boldsymbol{G}}_{\hat{E}|\hat{F}}\hat{\bmu}_{\hat{F}}\]
\[\hat{\bSigma}_{\hat{E}|\hat{F}}=\hat{\bSigma}_{\hat{E}}-\hat{\bSigma}_{\hat{E},\hat{F}}\hat{\bSigma}_{\hat{F}}^{-1}\hat{\bSigma}_{\hat{F},\hat{E}}\]
and update the value of the trimming function $\zeta(\cdot)$, setting $\zeta(\mathbf{x}_n)=0$ for the \(\lfloor N\gamma\rfloor \) samples with lowest value of
\[ \sum_{g=1}^G l_{ng} \log \left[ \hat{\tau}_g \phi(\mathbf{x}_{n,\hat{F}}; \hat{\boldsymbol{\mu}}_{g,\hat{F}}, \hat{\boldsymbol{\Sigma}}_{g,\hat{F}}) \right]
+\log \left[ \phi\left(\mathbf{x}_{n,\hat{E}}-\hat{\boldsymbol{G}}_{\hat{E}|\hat{F}}\mathbf{x}_{n,\hat{F}};\hat{\bmu}_{\hat{E}|\hat{F}},\hat{\bSigma}_{\hat{E}|\hat{F}}\right)\right]. \]
  \item Iterate $2-4$ until the $\lfloor N \gamma \rfloor$ discarded observations are exactly the same on two consecutive iterations, then stop.
\end{enumerate}
The procedure described in steps 1-5 shall be performed $\texttt{n\_init}$ times: the parameter estimates that lead to the highest value of the objective function \eqref{trim_ll_EMST_varsel}, out of $\texttt{n\_init}$ repetitions, provide the final estimated quantities. As a last worthy comment, notice that the specification of the cardinality of $F$, i.e., the number $p$ of relevant variables that are sought by the algorithm, is a-priori required as a model hyper-parameter. 

\subsection{Methods comparison} \label{sec:SRUW_vs_EMST}
In the previous subsections two novel methods for robust variable selection in model-based classification have been introduced. As already anticipated, the main operational difference between the two relies on the fact that the ML subset selector requires the a-priori specification of the subset-size $p$, whereas the greedy-forward approach via TBIC automatically infers the number of relevant variables by means of a stopping criterion in the stepwise search. This could come both as an advantage and as a disadvantage: one may desire to specifically retain the $p$ most relevant variables (i.e., $p=2$ for visualization purposes). In this case, the ML subset selector approach shall be preferred, as the entire feature space $P$ is accounted for in the likelihood specification in \eqref{trim_ll_EMST_varsel}, contrarily to the greedy approach employed in Section \ref{sec:SRUW}. If this is not the case, run the algorithm for a reasonable range of values $p$ and select the favourite solution, consensus methods like the one in \cite{strehl2002cluster} for clustering can be adapted to the classification framework. In addition, if computational burden is not an issue, the greedy-forward approach via TBIC can be firstly employed for assessing the order of magnitude of the subset size, and afterwards the ML subset selector can be run varying $p$ in the proximity of the number of relevant variables found by the former method, qualitatively assessing the difference.

Clearly, the suggestions above are mostly heuristic
, a more formal treatment on how to compare and validate results from both procedures is still missing: this however goes beyond the scope of the present manuscript and it will be the object of future research.
\subsection{On the choice of the impartial trimming level $\gamma$} \label{sec:choosing_gamma}
Both methodologies introduced in the previous sections require the a-priori specification of the impartial trimming level $\gamma$. Sensibly setting such a hyper-parameter is not an easy task, as it implicitly requires the user to estimate the degree of contamination he/she is expected to find in a dataset. How to automatically infer, in a data-driven fashion, the true noise percentage present in the data is still an open issue in the robust clustering literature, even though several procedures have been recently proposed to mitigate and/or partially solve the problem: see for example \cite{Garcia-Escudero2011, Dotto2018, Cerioli2018a, Cerioli2019, Riani2019a} and references therein. Given that, providing an exhaustive solution to this age-old problem goes beyond the scope of the present manuscript; nevertheless, some remarks and suggestions must be made on this regard.

Start by noticing that robust variable selection adds a layer of difficulty to the learning framework. More precisely, in this context we would firstly like to prevent adulterated units from jeopardizing the identification of relevant features, and, subsequently, to perform robust estimation and outlier detection on the subset of retained variables. While the two stages are clearly interconnected, 
we have observed that the performance of the former is generally less dependent by the choice of $\gamma$ than the latter. Particularly, numerical experiments have highlighted that underestimating the true contamination level tends to favor the inclusion of several redundant and/or irrelevant variables among the relevant ones (see Section \ref{sec:sim_study_varsel}). On the other hand, the sensitivity study reported in Section \ref{sec:sens_study_gamma} reveals that setting a higher than necessary trimming level has much less negative impact in the important features identification. Motivated by these arguments we suggest to, at least initially, set a precautionary high value of $\gamma$ when it comes to variable selection, and, once the relevant subset has been identified, to tune it on the so-obtained lower dimensional feature space by exploiting the techniques already available in the literature. Operationally on the other hand, an adaptive procedure can be built by starting the search from a high level of trimming $\gamma_{MAX}$, and subsequently monitoring the changes in the retained variables subset for decreasing values of $\gamma$. This solution stems from the ideas discussed in \cite{Riani2019a}, where the partition stability of a robust clustering procedure is monitored by computing the Adjusted Rand Index \citep[ARI;][]{Rand1971} between consecutive allocations. In our context, whenever the magnitude and/or the configuration of two consecutive subsets of retained variables change, it can be interpreted as a sign that some contaminated units could have spoiled the procedure. Notice that standard information theoretic concepts, like the Hamming distance \citep{hamming1950error}, can be directly used to compare solutions obtained via the ML subset selector method: the relevant subset size $p$ is a-priori set and assumed to be kept fixed during the search. On the other hand, more involved metrics, like, for example, Levenshtein distances \citep{dan1997algorithms}, are needed to compare outputs from the stepwise approach via TBIC, as their magnitude may vary when different trimming values are considered. 

As a last worthy note, we highlight the fact that our procedures are specifically designed to prevent attribute and class noise to deteriorate the variable selection process. Once this has been performed, extra effort shall be invested in effectively unraveling how many anomalous observations are contaminated within the set of retained features. That is, we are not directly interested in identifying outliers that can certainly arise in those dimensions that are not important for the classification task: a relevant example on this matter is showcased by sample $43$ on the starches discrimination study in Section \ref{sec:application_varsel}. To this extent, coupling cellwise outlier detection \citep{Farcomeni2014, Rousseeuw2018} and variable selection could be a promising attempt to cast light on this problem: it will be the object of future research.


\section{Simulation studies} \label{sec:sim_study_varsel}
The aim of this Section is to numerically assess the effectiveness of the methodologies introduced in Section \ref{sec:robust_var_sel}, whilst investigating the effect that contamination has on standard variable selection procedures. In doing so, we decided to rely on the same data generating process (DGP) considered in \cite{Maugis2011} and \cite{Celeux2019}, described in Section \ref{sim_setup}, including in addition some attribute and class noise to the original experiment. Firstly, a simulated example with a fixed level of contamination is presented in Section \ref{sec:sim_study_gamma_fixed}, in which accuracy metrics are computed for both robust and non-robust variable selection methods. Secondly, a sensitivity study is reported in Section \ref{sec:sens_study_gamma}, displaying how different trimming and contamination levels affect the novel procedures.

\subsection{Experimental Setup} \label{sim_setup}
The synthetic dataset considers $G=4$ classes for a total of $P=16$ features: the first three are relevant for the classification, the subsequent four are redundant given the first ones, while the last nine are independent from both the group variable and the previous predictors. The prior probabilities of the four classes are equal to $\boldsymbol{\tau}=(0.15, 0.3, 0.2, 0.35)$. On the three discriminant variables, data are generated from multivariate normal densities
\begin{equation*}
\mathbf{x}_n^{[1-3]}|\mathbf{l}_n=g \sim \phi\left(\boldsymbol{\mu}_g, \boldsymbol{\Sigma}_g\right), \:\:\:\:g=1,\ldots,4
\end{equation*}
with mean vectors
\[ \boldsymbol{\mu}_1=(1.5, -1.5,1.5)', \quad \boldsymbol{\mu}_2=(-1.5, 1.5,1.5)'\]
\[\boldsymbol{\mu}_3=(1.5, -1.5,-1.5)',\quad \boldsymbol{\mu}_4=(-1.5, 1.5,-1.5)' \]
and covariance matrices $\boldsymbol{\Sigma}_g$ with elements $\rho_g^{|i-j|}$, $1 \leq i,j \leq 3$, and $\rho_1=0.85$, $\rho_2=0.1$, $\rho_3=0.65$, $\rho_4=0.5$. The four redundant variables are sampled from
\[\mathbf{x}_n^{[4-7]}\sim N\left(\mathbf{x}_n^{[1,3]}\boldsymbol{B}; \mathbf{I}_4 \right), \quad \boldsymbol{B}=\begin{pmatrix}
  1 & 0 & -1 &0\\
  0 & -2 & 2 & 1
  \end{pmatrix},
\]
while the 9 independent ones are simulated from $\mathbf{x}_n^{[8-16]}\sim N(\boldsymbol{\eta}, \boldsymbol{\delta})$ with
\[\boldsymbol{\eta}=(-2, -1.5, -1, -0.5, 0, 0.5, 1, 1.5, 2)\]
and
\[\boldsymbol{\delta}=\mbox{diag}(0.5, 0.75, 1, 1.25, 1.5, 1.25, 1, 0.75, 0.5).\]
A total of $B=100$ Monte Carlo (MC) samples are produced in both the simulation and the sensitivity study. Results are reported in the next Sections.

\subsection{Simulation experiment with fixed level of contamination} \label{sec:sim_study_gamma_fixed}
From the DGP outlined in Section \ref{sim_setup}, $N=500$ units are generated and their group membership retained for constructing the training set; while $M=5000$ unlabeled observations compose the test set. Subsequently, label noise is simulated by wrongly assigning $20$ units coming from the fourth group to the third class. In addition, $5$ uniformly distributed outliers, having
\begin{itemize}
\item squared Mahalanobis distances $d\left(\cdot; \boldsymbol{\mu}_g, \boldsymbol{\Sigma}_g\right)$ greater than $\chi^2_{3, 0.975}$ $\forall g \in\{1,2,3,4\}$ for the relevant variables
\item $d\left(\cdot; \boldsymbol{\mu}_g^{[1,3]}B, \boldsymbol{I}_4 \right)$ greater than $\chi^2_{4, 0.975}$ $\forall g \in\{1,2,3,4\}$ for the redundant variables
\item $d\left(\cdot;\boldsymbol{\eta},\boldsymbol{\delta} \right)$ greater than $\chi^2_{9, 0.975}$ for the irrelevant variables
\end{itemize}
are appended to the training set, with randomly assigned labels. This contamination produce, in each MC replication, a total of $25$ adulterated units, that account for slightly less than $5\%$ of the entire learning set. We validate the performance of our novel methods in correctly retrieving the relevant variables, compared to non-robust procedures. Particularly, the comparison is carried out considering the following methods:
\begin{itemize}
\item TBIC: robust stepwise greedy-forward approach via TBIC (Section \ref{sec:SRUW})
\item ML subset: maximum likelihood subset selector approach (Section \ref{sec:EMST}), with subset size of relevant variables $p$ equal to $3$, $6$ and $9$
\item SRUW: stepwise greedy-forward approach via BIC \citep{Maugis2011}
\item SelvarMix: variable selection in model-based discriminant analysis with a regularization approach \citep{Celeux2019}.
\end{itemize}
Furthermore, once the important variables have been identified, the associated classifier (i.e., REDDA for the robust variable selection criteria and EDDA for the non-robust ones) is trained on the reduced set of predictors and the classification accuracy is computed on the test set. A labeled trimming level $\gamma$ equal to $0.05$ was kept fixed during the experiment. Lastly, for providing benchmark values on the relevance of feature selection, both EDDA and REDDA classifiers are also fitted on the original set with $P=16$ variables.

\begin{figure}
\includegraphics[scale=.58]{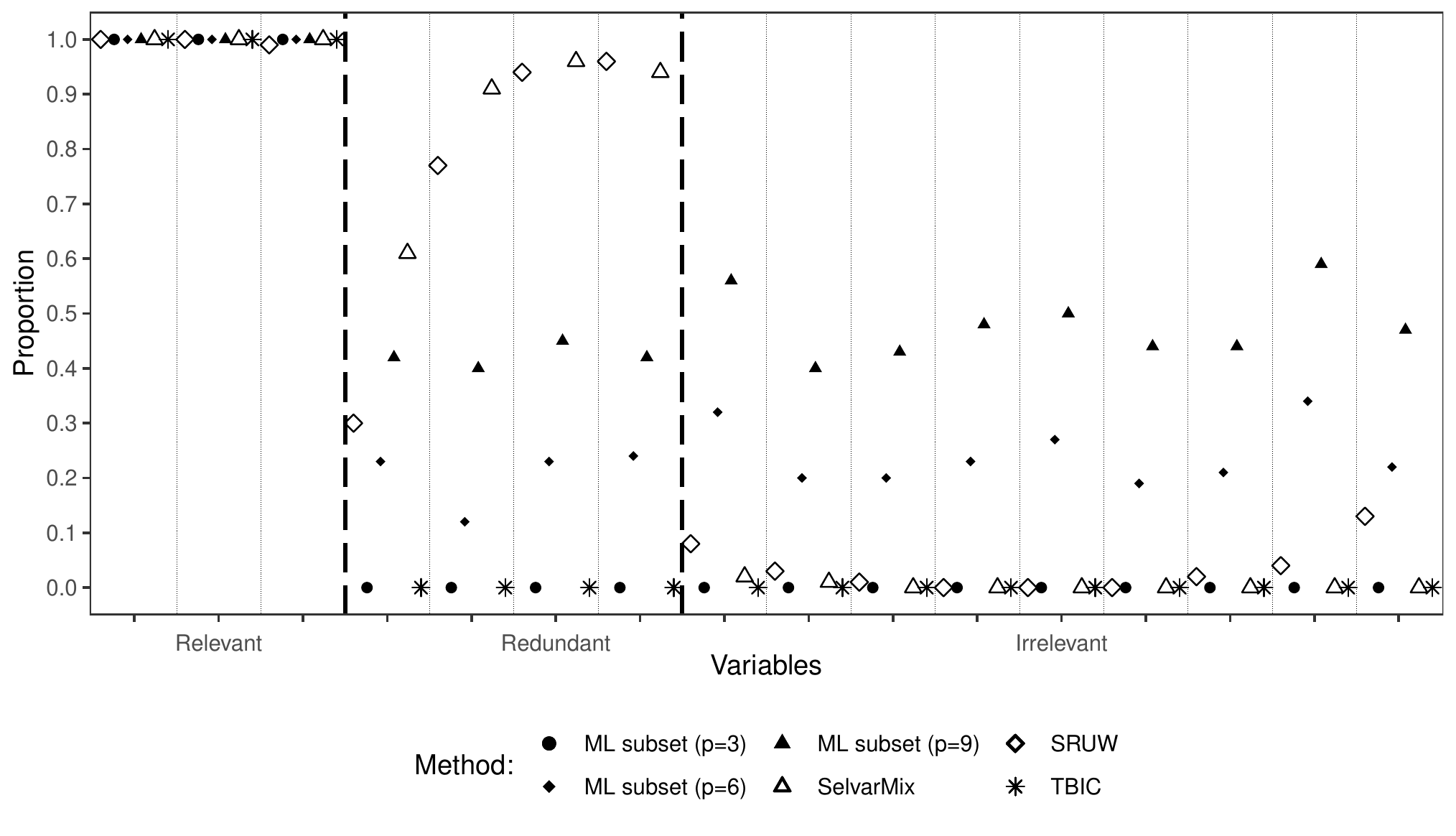}
\centering
\caption{Proportion of times a variable has been
selected as relevant, out of $B=100$ MC repetition of the simulated experiment, for different variable selection methods.}
\label{fig:sim_study_varsel_proportion}
 \end{figure}
Figure \ref{fig:sim_study_varsel_proportion} displays the proportion of times a variable has been selected as relevant by the different methods in the $B=100$ repetitions of the simulated experiment. As it is clearly visible from the plot, the first three features are selected by all the procedures in almost every iteration of the simulation study. 
Generally, therefore, the contamination introduced in the training set does not cause any systematic exclusion of 
the true discriminative variables from the relevant subset, also for the non-robust methods.
\begin{figure}
\includegraphics[scale=.58]{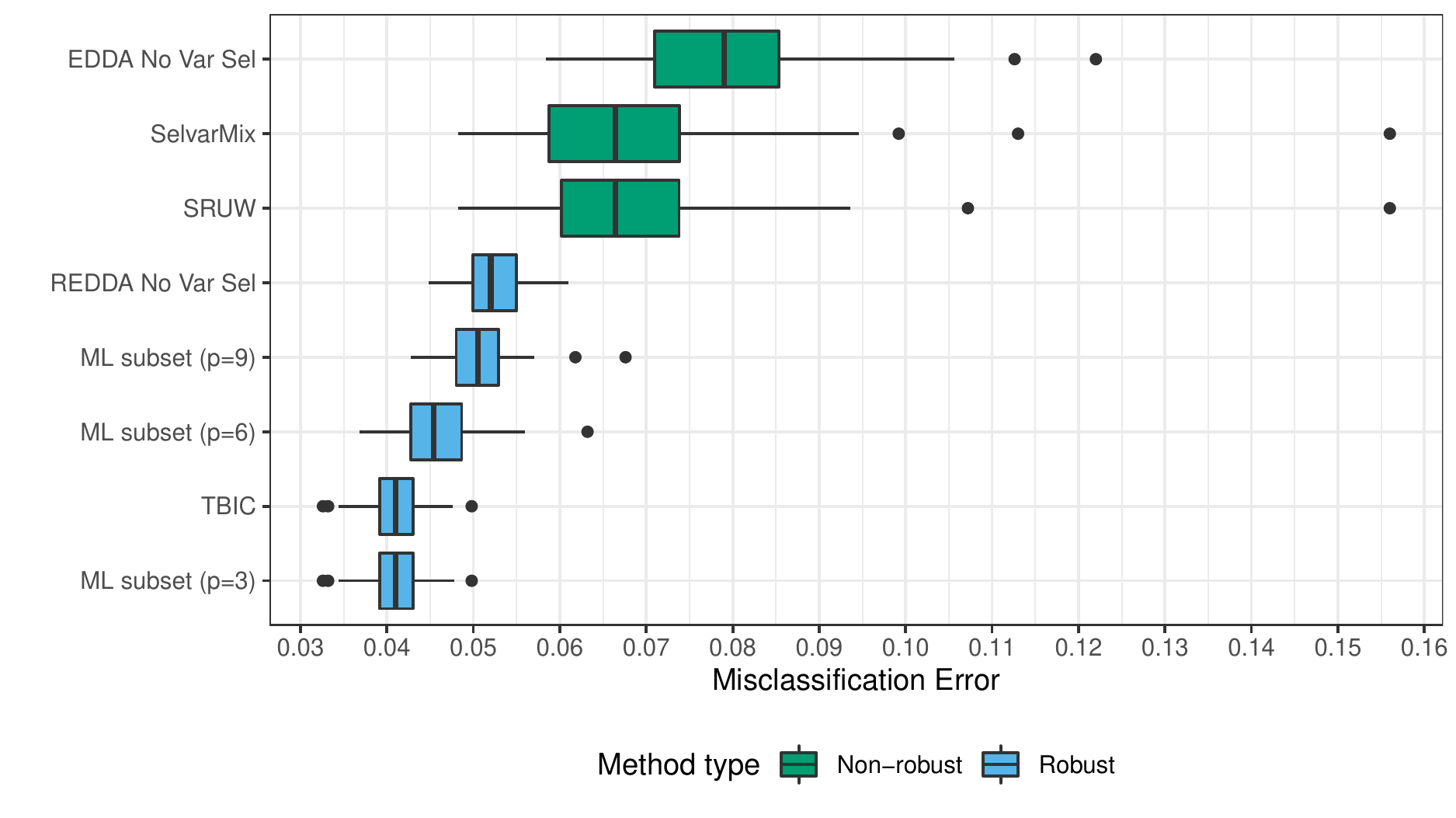}
\centering
\caption{Boxplots of the misclassification error, out of $B=100$ MC repetition of the simulated experiment, for the $M=5000$ test data, varying variable selection and model-based classification methods.}
\label{fig:box_plot_misclass_error_sim_study}
 \end{figure}
Nonetheless, outliers and label noise lead SRUW and SelvarMix to severely overestimate the number of retained features. Redundant and irrelevant variables are often included in the selection, as demonstrated by the hollow triangles and diamonds in Figure \ref{fig:sim_study_varsel_proportion}. The robust stepwise approach via TBIC, instead, does not seem to suffer from this unfavorable behavior: it correctly identifies the first three relevant variables in every single simulation. As already pointed out in Section \ref{sec:SRUW_vs_EMST}, the main drawback of the maximum likelihood subset selector approach is given by the need of pre-specifying the subset size $p$. When $p=3$, i.e., the true number of discriminating variables, the algorithm always correctly selects the relevant ones. Clearly, when $p$ is set higher than three, some irrelevant and/or redundant features will be necessarily included in the retained set. However, letting $p$ to be greater than the true relevant predictors does not seem to severely affect the predictive power of the robust classification rule. As it can be seen from the results reported in Table \ref{tab:misclass_error_sim_study} and in Figure \ref{fig:box_plot_misclass_error_sim_study}, the misclassification errors are only slightly influenced by the choice of $p$ in the ML subset selector, and are always on average lower than non-robust procedures. As expected, the best prediction accuracy is obtained when $p=3$, result that entirely agrees with the one obtained by the forward selection algorithm via TBIC, as the very same variables are selected for each simulation and, subsequently, the REDDA classifier is fitted on the retained subset.
\begin{table}[ht]
\centering
\caption{Average misclassification error, out of $B=100$ MC repetition of the simulated experiment, for the $M=5000$ test data, varying variable selection and model-based classification methods. Standard deviations reported in parentheses.}
\label{tab:misclass_error_sim_study}
\centering
\begin{tabular}{rlrl}
  \hline
Method & Misclassification Error & Method & Misclassification Error\\
  \hline
ML subset (p=3) & 0.0411 & REDDA No Var Sel & 0.0525 \\ 
   & (0.003) &  & (0.003) \\ 
  ML subset (p=6) & 0.0457 & SRUW & 0.0686 \\ 
   & (0.0045) &  & (0.0045) \\ 
  ML subset (p=9) & 0.0506 & SelvarMix & 0.0684 \\ 
   & (0.004) &  & (0.004) \\ 
  TBIC & 0.0411 & EDDA No Var Sel & 0.0795 \\ 
   & (0.003) &  & (0.003) \\ 
   \hline
\end{tabular}
\end{table}
Interestingly, the EDDA classifier coupled with (non-robust) variable selection via either SelvarMix or SRUW shows on average higher misclassification error than REDDA learned on the entire set of features. That is, the harmful effect of adulterated observations is increased by the presence of noisy variables, also shown by the poor performance of EDDA with no feature selection. The present simulation study highlights how a very small proportion of attribute and class noise may somewhat spoil a wrapper procedure, driving the algorithm to include many more features than the truly relevant ones. That is, when adulterated units are not properly dealt with, both feature identification and classification may provide inappropriate results, with bias in the former propagating to badly affect  the derived classifier even further. 

\subsection{Sensitivity study varying contamination and trimming levels} \label{sec:sens_study_gamma}
\begin{figure}
\includegraphics[scale=.5]{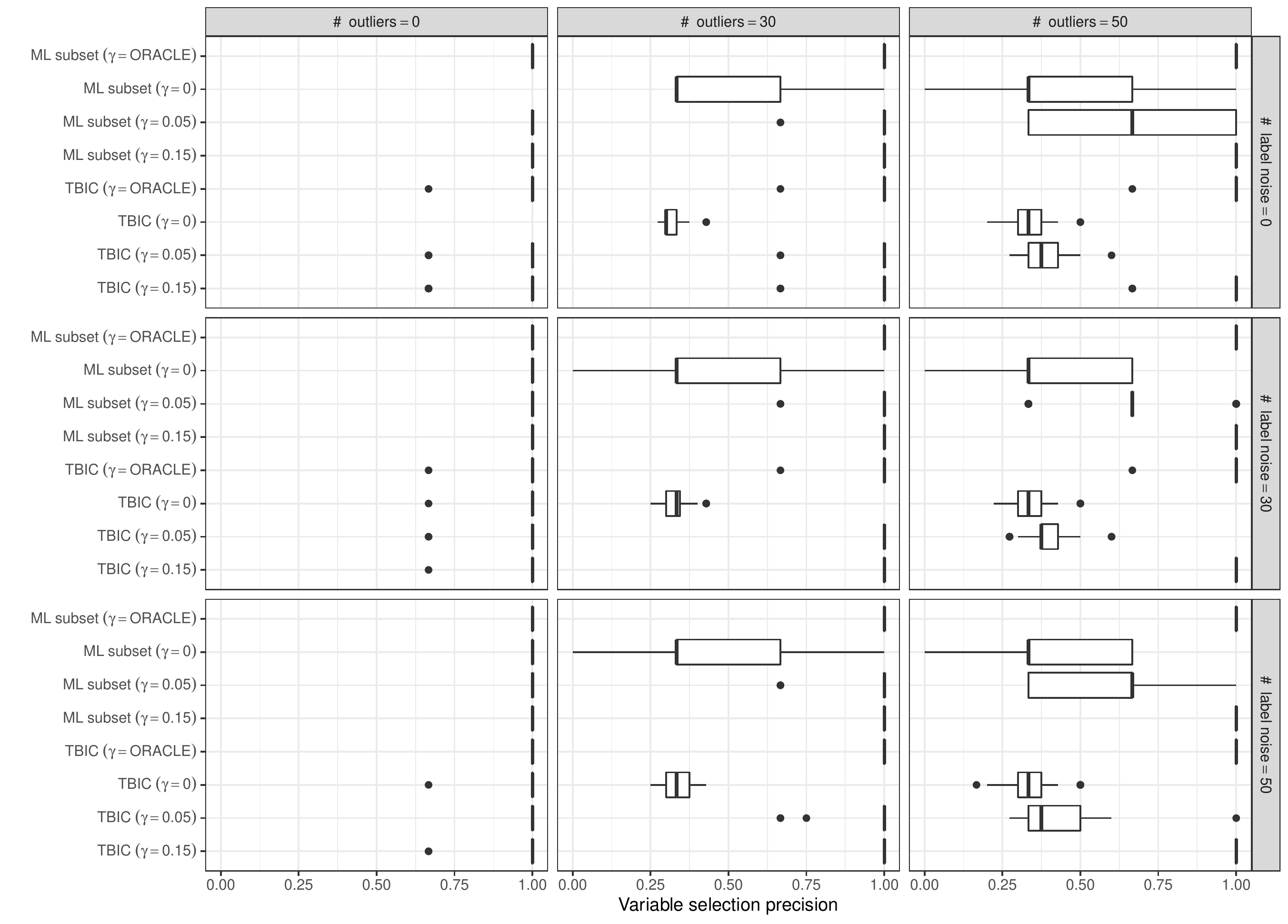}
\centering
\caption{Boxplots of variable selection precision, out of $B=100$ MC repetition of the sensitivity study, varying contamination and trimming levels. The boxplots of TBIC $(\gamma=0)$ and ML subset $(\gamma=0)$ for the uncontaminated scenario (top left corner) are not displayed since they agree with the oracle solution. For the ML subset selector approach, the subset size of relevant variables $p$ is set equal to $3$.}
\label{fig:boxplot_sens_study}
 \end{figure}

A sensitivity study is built upon the same DGP outlined in the previous Sections, encompassing $9$ different scenarios varying the true proportions of attribute and class noise in the training set. The actual number of mislabeled and outlying units ,and resultant contamination rate are reported in Table \ref{tab:sens_study_table}. The aim of this experiment is to numerically investigate the effect that the misspecification of $\gamma$, i.e., to overestimate or to underestimate the true contamination level, produces in the novel variable selection procedures.
\begin{table}[ht]
\centering
\caption{Number of outliers, label noise and resultant true contamination rate for the $9$ scenarios encompassing the sensitivity study.}
\label{tab:sens_study_table}
\begin{tabular}{lrrrrrrrrr}
  \hline
\# Outliers & 0 & 0 & 0 & 30 & 30 & 30 & 50 & 50 & 50 \\ 
\# Label Noise & 0 & 30 & 50 & 0 & 30 & 50 & 0 & 30 & 50 \\ 
   \hline
Contamination  & 0 & 0.06 & 0.10 & 0.057 & 0.113 & 0.151 & 0.091 & 0.145 & 0.182 \\ 
   \hline
\end{tabular}
\end{table}
The results obtained in Section \ref{sec:sim_study_gamma_fixed} display a general tendency in overstating the important subset magnitude for non-robust solutions. Motivated by this fact, we are most interested in evaluating the ability of our methods in solely retrieving the $3$ relevant features. To this extent, we employ the ``variable selection precision'' metric as a measure of performance, defined to be the proportion of features selected by a methodology truly belonging to the relevant subset, out of all the retained ones. For each scenario reported in Table \ref{tab:sens_study_table}, we fit both the robust stepwise via TBIC and the ML subset selector (with $p$ fixed and equal to $3$) approaches considering impartial trimming levels $\gamma$ respectively equal to $0$, $0.05$ and $0.15$. In addition, we also evaluate the precision of the so-called ``oracle'' solutions, obtained by setting $\gamma$ equal to the true contamination rate. Sensitivity study results are summarized in Figure \ref{fig:boxplot_sens_study}, in which several interesting patterns emerge.
 
First off, it is immediately noticed that the overall precision is much more badly affected by outliers addition rather than by mislabeled samples. Such a behavior is explained by recognizing that label noise does not directly alter the feature space: classes separation gets indeed more uncertain, yet the most discriminative variables are still identified even when $\gamma$ is underestimated. The same does not happen when it comes to attribute noise; as soon as the number of outliers exceeds the trimming level the obtained precision deteriorates for both methodologies. Notwithstanding, solutions based on high values of $\gamma$ (i.e., $\gamma=0.15$) adequately safeguard the procedure performance, even in the most extreme scenario (for which the true contamination level is equal to $0.182$).

Secondly, we observe that, conditioning on the same $\gamma$ value, the approach based on stepwise TBIC generally shows worse performances with respect to the ML subset selector method. Needless to say, the fact of a-priori setting the number of retained variables $p$ equal to the true relevant subset has a major positive impact for the latter method. In details, the stepwise TBIC procedure tends to include unnecessary features when $\gamma$ is underestimated, with consequent loss in variable selection precision. On the other hand, the ML subset selector is forced to select only the $p=3$ most relevant ones. In spite of that, for the scenarios in which the outliers proportion exceeds the trimming level we notice a reduction in terms of estimated precision: the method fails in distinguishing between redundant and relevant features.

Lastly, and perhaps most importantly, the results highlight that cautiously overestimating the contamination level does not impact the ability of our methods in retrieving the relevant subset; the same cannot be said when $\gamma$ is set lower than needed. The ``trimming more is better than trimming less''  principle is well known for robust methods based on hard-trimming, it seems however particularly true for our variable selection procedures, for which no corresponding drawback to the efficiency loss in parameter estimation has been identified in our synthetic experiments. Clearly, further considerations are needed to formally assess this promising property, and, to this extent, a separate line of research is currently being pursued.


All in all even though, as already pointed out in Section \ref{sec:choosing_gamma}, properly choosing $\gamma$ still remains a critical step in all robust procedures based on impartial trimming, this sensitivity study underlines how variable selection seems not to be badly affected by an overestimation of the contamination level. Therefore, replacing standard methods with robust solutions seems paramount whenever it is believed the considered dataset may contain some noisy units, especially in high dimensional settings.

 
\section{Application to MIR spectra: starches discrimination} \label{sec:application_varsel}
Chemometrics is a natural field of application for high-dimensional statistics, as data recorded from chemical systems are complex in nature and generally limited in terms of sample size. In particular, variable selection methods are notably appealing for observations recorded by spectroscopic instruments: for virtually continuous spectra
the information contained in adjacent features is often correlated, and thus the determination of a relevant subset of wavelengths is desirable, prior to perform any subsequent analysis \citep{Brown1992, Brenchley1997}. Furthermore, data reduction simplifies results interpretation, making future measurements simpler and cheaper \citep{Indahl2004}.

Spectroscopic data are recorded during a controlled experiment, and the quality of both measurements and analysed substances is, in most cases, reliable. Nevertheless, calibration errors may appear during spectra collection, and, moreover, for some delicate applications such as food authenticity, the raw material itself may be spoiled and/or adulterated \citep{Reid2006}. In this context, therefore, variable selection methods that not only robustly identify relevant wavelengths, but also recognize outliers and possibly fraudulent samples may be particularly valuable to chemometricians. Motivated by a Mid-infrared (MIR) dataset of the chemometrics challenge organized during the `Chimiom\'{e}trie 2005' conference, the methodologies introduced in Section \ref{sec:robust_var_sel} are employed for performing high-dimensional classification and outlier detection.

\subsection{Data} \label{data_application_varsel}

\begin{figure}
\includegraphics[scale=.55]{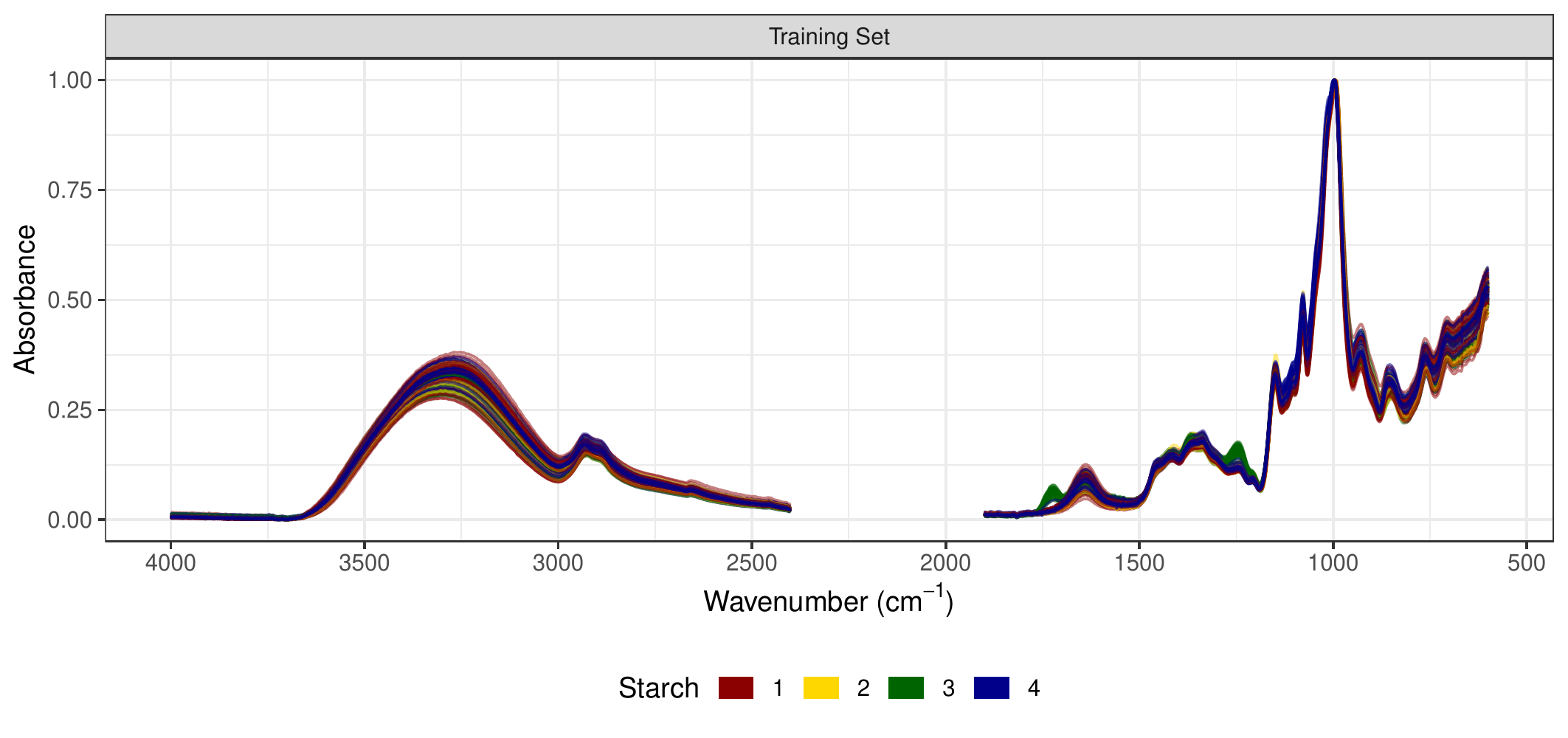}
\centering
\caption{Mid infrared spectra of starches of four different classes, training set.}
\label{fig:starches_training}
 \end{figure}

The considered datasets, described in \cite{FernandezPierna2005, FernandezPierna2007a}, include respectively $N=215$ (training set) and $M=43$ (test set) MIR spectra of starches of four different classes, taken on a Perkin-Elmer Spectrum $2000$ FTIR spectrometer (Perkin Elmer Corporation, Norwalk, CT, USA) between $4000$ and $600$ $cm^{-1}$ at $1$ $cm^{-1}$ data interval. The range between $2402$ and $1901$ $cm^{-1}$ was removed from the spectra, so that a total of $P=2901$ absorbance measurements were then retained for the analysis. A subset of the learning observations is displayed in Figure \ref{fig:starches_training}.
In order to create an extra difficulty to be tackled by the participants during the competition, four outliers were included in the test set:
\begin{itemize}
\item \textit{Sample 2:} a shifted version of unit 1, obtained by removing its first six data points and appending six new variables at the end of the spectrum;
\item \textit{Sample 4:} a noisy version of unit 2, by generating Gaussian white noise and adding it to the absorbance values of the sample;
\item \textit{Sample 43:} a modified version of unit 39, obtained by manually changing a data point on the spectrum (wavelength $2456$) to simulate a spike;
\item \textit{Sample 20:} a modified version of unit 17, by adding a slope to its original spectrum.
\end{itemize}
Therefore, the discrimination challenge held during `Chimiom\'{e}trie 2005' consisted in learning a classification rule from the training set to predict the labels of the test units, whilst also performing adulteration detection on the latter. In our experiment, we additionally include label noise by  wrongly assigning the last four units of the third group of starches to the fourth one: this accounts for less than $2\%$ of the entire training set. Classification results are reported in the next Section.
 
\subsection{Results}
\begin{figure}
\includegraphics[scale=.4]{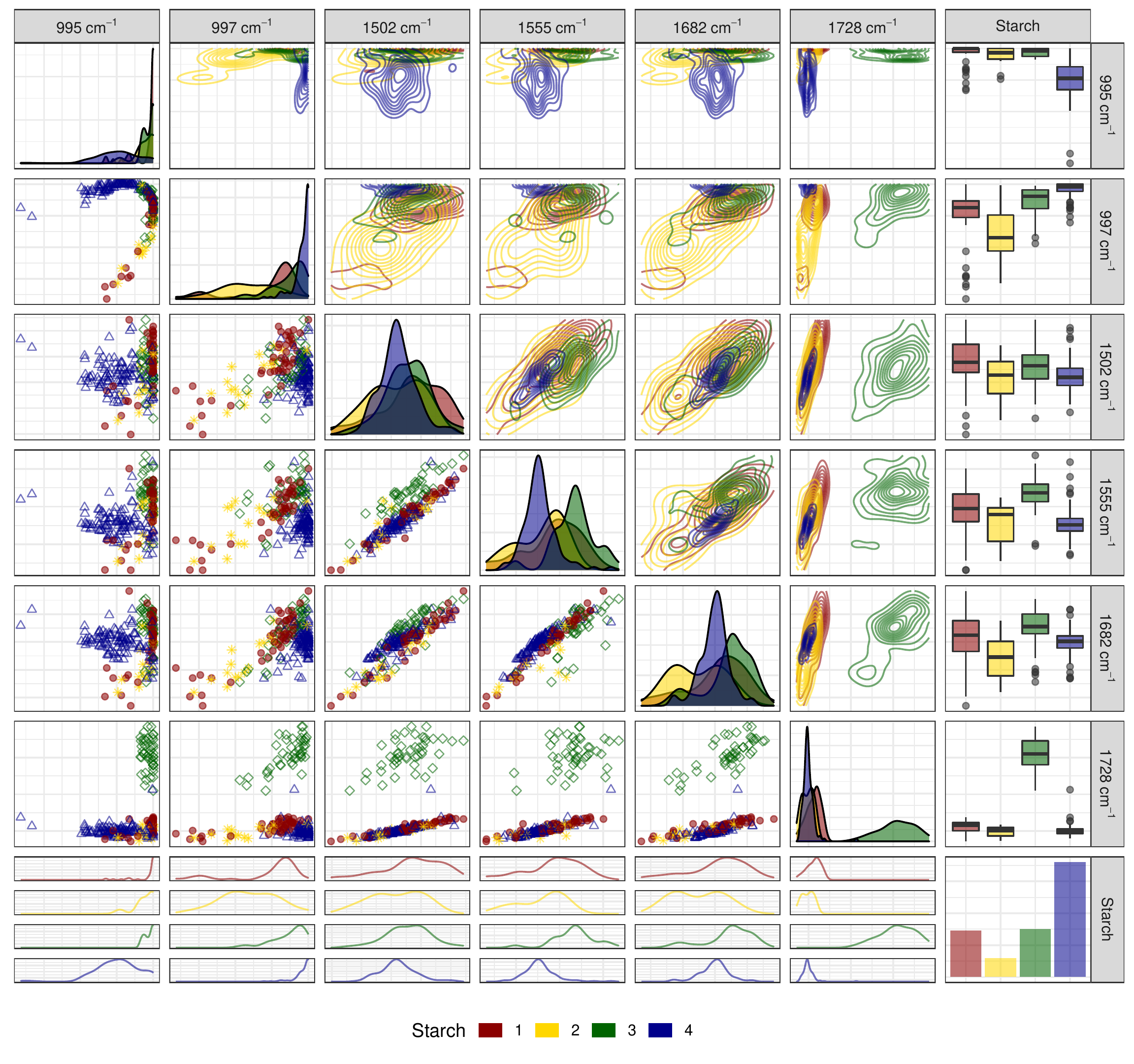}
\centering
\caption{Generalized pairs plot of the relevant variables selected by the stepwise greedy-forward approach via TBIC.  Starches dataset, training samples.}
\label{fig:pair_plot_rel_var_TBIC}
 \end{figure}

\begin{figure}
\includegraphics[scale=.5]{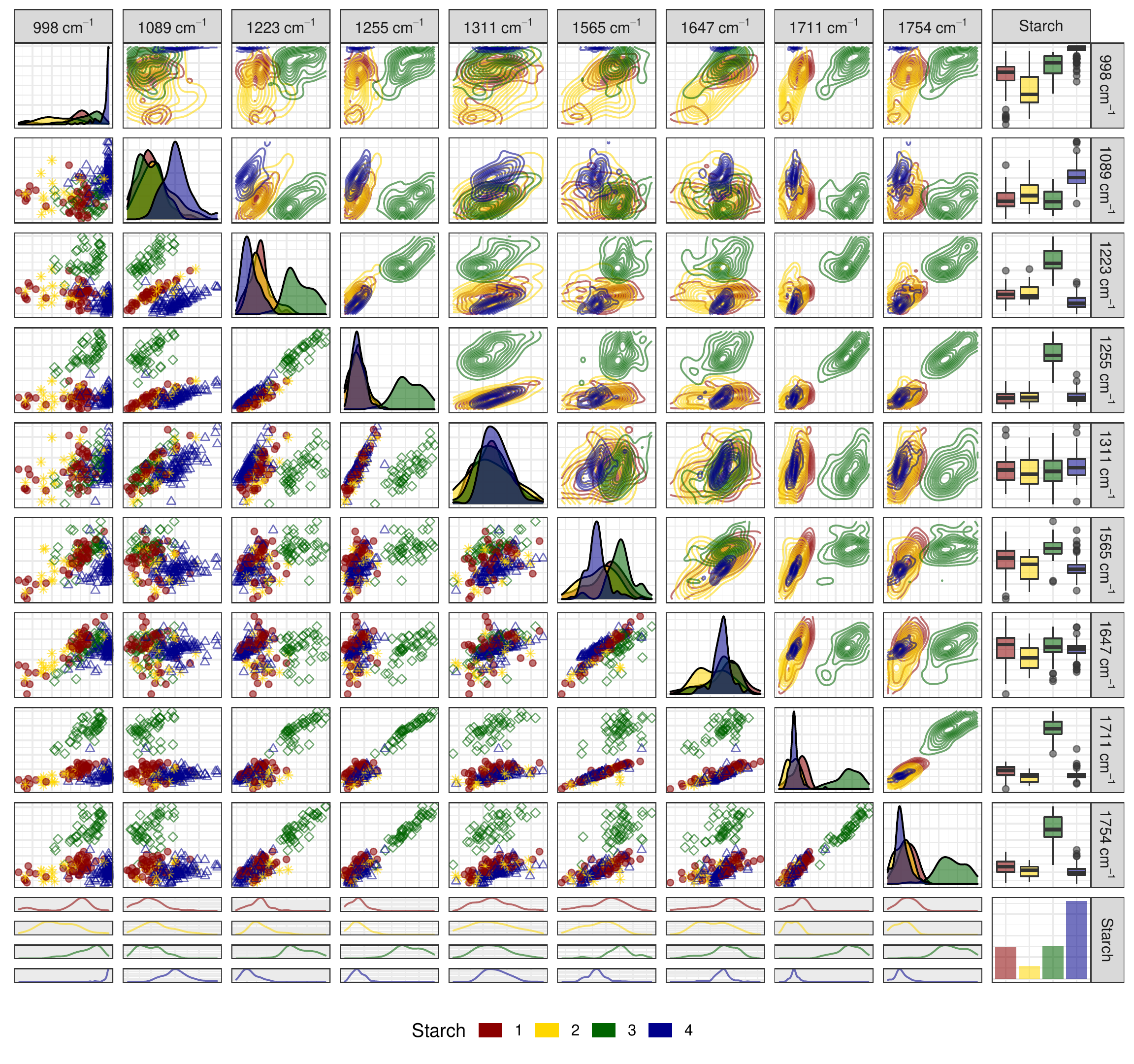}
\centering
\caption{Generalized pairs plot of the relevant variables selected by ML subset selector with $p=9$.  Starches dataset, training samples.}
\label{fig:pair_plot_rel_var_EMST}
 \end{figure}

The discriminating problem described in the previous Section cannot be solved by directly applying model-based classifiers, since $N \ll P$. To overcome this issue, we make use of the robust wrapper variable selection methods introduced in this article: such approaches provide a natural solution for dealing with contaminated high-dimensional data, and, as we will see, they can be further used to identify the noisy units in the test set. We firstly run the stepwise greedy-forward approach via TBIC (Section \ref{sec:SRUW}) with $\gamma=0.05$: the procedure, out of $P=2901$, selects a total of only six relevant wavelengths: $1728$ cm$^{-1}$, $1682$ cm$^{-1}$, $1555$ cm$^{-1}$, $1502$ cm$^{-1}$, $997$ cm$^{-1}$ and $995$ cm$^{-1}$. Figure \ref{fig:pair_plot_rel_var_TBIC} displays the generalized pairs plot for the selected variables.
Motivated by the TBIC output and by the results presented in the Simulation Study, 
we decided to retain a slightly higher number of relevant variables in the ML subset selector, setting the value of  $p$ to be equal to $9$. In doing so, the ML subset selector estimates the relevant subset $F$ to be comprised of the following wavelengths: $998$ cm$^{-1}$, $1089$ cm$^{-1}$, $1223$ cm$^{-1}$, $1255$ cm$^{-1}$, $1311$ cm$^{-1}$, $1565$ cm$^{-1}$, $1647$ cm$^{-1}$, $1711$ cm$^{-1}$ and $1754$ cm$^{-1}$. A generalized pairs plot \citep{Emerson2013} of such subset is reported in Figure \ref{fig:pair_plot_rel_var_EMST}.
Interestingly, the two approaches select entirely different wavelengths as the most discriminative ones. Careful investigation of this behavior shows high correlation between the variables selected by the two methodologies, while the correlation reported by features within the same subset is much lower. Clearly, in dealing with real datasets the separation between relevant, irrelevant and redundant variables is much less apparent. Particularly for spectroscopic data, highly correlated wavelengths often result in comparable discriminating power, with no natural preference in terms of relevance. Nevertheless, it is worth noting that both methods chose wavelengths from the right-hand side part of the spectrum, as it seems to delineate the highest separation between the different starches, also by visual inspection of Figure \ref{fig:starches_training}.

A REDDA model with $\gamma=0.05$ is employed to predict the class for the test samples, using as predictors the variables retained by the TBIC and ML subset selector, respectively. In both cases, units that present class noise in the training set were correctly identified as such and not accounted for in the estimation procedure. In addition, a Support Vector Machine with Gaussian radial kernel (SVM) was also considered, as it was shown to be the best performing classifier for this specific dataset \citep{FernandezPierna2005,FernandezPierna2007a}. Lastly, we replicate the second best solution proposed by one of the contest participants: an ensemble method was constructed by combining ROC, PLS and SVM predictions via majority vote on a subset of variables, previously determined by a PLS model. Classification accuracy for the four competing methods, considering test sets without modified units, is reported in Table \ref{tab:misclass_error_application}: 

\begin{table}[ht]

\centering
\caption{Number of correctly predicted test samples and associated misclassification error for different methods. The test set without outliers has a total sample size of $M=39$.}
\label{tab:misclass_error_application}
\begin{tabular}{rrrrr}
  \hline
 & REDDA& REDDA & SVM & ROC+PLS+SVM\\ 
 &(TBIC)& (ML subset)& radial kernel &\\ 
  \hline
 $\#$correctly predicted & 32 & 34 & 31 &31\\ 
    Misclassification error & 0.179 & 0.128 & 0.205 & 0.205\\ 
   \hline
\end{tabular}
\end{table}
the robust model-based classifiers show better results than the other solutions. The performance of the kernel and ensemble methods are negatively impacted by the presence of the 4 mislabelled units in the training set: compare results in Table \ref{tab:misclass_error_application} with the ones reported in Table 1 of \cite{FernandezPierna2007a}, wherein the classifiers were trained on an uncontaminated learning set. The relevant subsets retained by both robust variable selection methods lead to similar results in terms of classification accuracy, with a slight better performance when REDDA is fitted on the features identified by the ML subset selector approach. As already pointed out in \cite{FernandezPierna2007a}, the main source of error is due to the difficulties in separating classes 1 and 2, as it is evident also in Figures \ref{fig:pair_plot_rel_var_TBIC} and \ref{fig:pair_plot_rel_var_EMST}.

We mentioned at the beginning of the Section that the REDDA method can be effectively employed in performing outlier detection in the test set. Particularly, given the probabilistic assumptions that underlie the methodology, for each test unit $\mathbf{y}_m$, $m=1,\ldots,M$, we can compute its estimated marginal density as follows:
\begin{equation} \label{d_value_varsel}
\hat{p}(\mathbf{y}_{m,\hat{F}};\hat{\tau},\hat{\boldsymbol{\mu}}_{\hat{F}}, \hat{\boldsymbol{\Sigma}}_{\hat{F}})=\sum_{g=1}^G \hat{\tau}_g \phi \left(\mathbf{y}_{m,\hat{F}}; \hat{\boldsymbol{\mu}}_{g,\hat{F}}, \hat{\boldsymbol{\Sigma}}_{g,\hat{F}} \right)
\end{equation}
where $\hat{F}$ denotes either the relevant variables identified by the stepwise approach with TBIC or by the ML subset selector, with parameters robustly estimated via the REDDA model on the retained features. For both variable selection approaches, the $3$ observations $\mathbf{y}_m $ with lowest value of \eqref{d_value_varsel} are units $2$, $4$ and $20$; all of them were manually modified, as described in Section \ref{data_application_varsel}. The only neglected outlier is unit $43$: it was contaminated on a single wavelength that was not identified as relevant by the variable selection methods. Nonetheless, by using an impartial trimming approach, we are effectively able to identify 3 out of 4 adulterated units.

In this Section, we have shown that the proposed noise-resistant variable selection approaches, coupled with robust discriminant analysis, can be effectively employed in performing high-dimensional classification in an adulterated framework. Even though being notably noise tolerant, powerful classifiers such as Support Vector Machine provide lower classification accuracy when a small percentage of class noise is present in the training set. In addition, after parameters have been robustly estimated, our proposal can be used to recognize possible adulterated units in the test set. All in all, an automatic methodology that performs robust feature detection, parameter estimation and outlier identification may become beneficial in chemometrics, easing both pre and post processing steps of complex spectroscopic analyses. 
 
\section{Concluding Remarks} \label{sec:conclusion_varsel}
In the present manuscript we have introduced two wrapper variable selection methods, resistant to outliers and label noise. We have shown that by means of these approaches we can effectively perform high-dimensional discrimination in an adulterated scenario. The first wrapper method embeds a robust model-based classifier within a greedy-forward algorithm, validating stepwise inclusion and exclusion of variables from the relevant subset via a robust information criterion. Some theoretical justifications that corroborates the procedure are also discussed. The second wrapper method resorts to the theory of maximum likelihood and irrelevance, defining an objective function in which the subset of relevant variables is regarded as a parameter to be estimated. A dedicated algorithm for MLE within a Gaussian family of patterned models has been developed, and practical implementation issues have been considered. Further, pros and cons of the two novel procedures have been discussed. The robust stepwise approach via TBIC enjoys the automatic identification of the relevant subset size, and it has displayed less variability in terms of selection precision. On the other hand, the ML subset selector is computationally faster and can be specifically useful in situations for which the number of important variables is known in advance, even though being this occasionally the case in applications. A simulation study has been developed for assessing the effectiveness of our proposals in recovering the true discriminative features in a contaminated scenario, comparing their performances against well-known variable selection criteria. The novel methods have then been successfully applied in solving a high-dimensional classification problem of contaminated spectroscopic data. High discriminating power has been exhibited by the final models, whence the identification of the wrongly labeled and/or adulterated observations is derived as a by-product of the estimation procedures.

An open point for further research regards the extension of the fully supervised framework outlined here to the adaptive one, where unobserved classes in the test set need also to be discovered, embedding the resulting semi-supervised procedure within a robust variable selection approach. 
In addition, careful investigation will be devoted to the development of a methodology that automatically assesses the contamination rate present in a sample, as the a-priori specification of the trimming level still remains an open issue in this field, particularly delicate for high-dimensional data.

\section*{Acknowledgments}
The authors are grateful to Prof. Dr. Gunter Ritter for the stimulating discussions and suggestions on how to transpose the ML subset selector approach, originally developed for clustering, to the classification framework. 
Thanks are due to Professor Ludovic Duponchel for providing relevant context on how the novel methodologies may be favorably employed, as well as for supplying the dataset. The authors also thank the editor and the two anonymous referees: their valuable comments greatly improved the quality of the paper. 
Brendan Murphy's work is supported by Science Foundation Ireland grants (SFI/12/RC/2289\_P2 and 16/RC/3835). Andrea Cappozzo and Francesca Greselin's work is supported by Milano-Bicocca University Fund for Scientific Research,  2019-ATE-0076.

\appendix
\section{Further aspects for the robust stepwise greedy-forward approach via TBIC} \label{sec:appendix_A} 
In this Section we retrieve the ML estimates for the grouping and no grouping structures in the robust stepwise greedy-forward approach (Section \ref{sec:SRUW}), by means of the spurious outliers model specification.
\subsection*{Grouping Model}
The log-likelihood function of the spurious outliers model under the grouping structure is:
\begin{align}
\begin{split}
\ell(\mathcal{N},\boldsymbol{\tau}^{cp}, \boldsymbol{\mu}^{cp}, \boldsymbol{\Sigma}^{cp})=&\sum_{n=1}^N q_n \sum_{g=1}^G l_{ng} \log{\left(\tau_g^{cp} \phi(\mathbf{x}_n^{c}, x_n^p; \boldsymbol{\mu}_g^{cp}, \boldsymbol{\Sigma}_g^{cp})\right)}+\\
&+\sum_{n=1}^N(1-q_n)\log{w(\mathbf{x}_n^{c}, x_n^p, \mathbf{l}_n;\boldsymbol{\psi}_n)}
\end{split}
\end{align}
to be maximized with respect to $\{\mathcal{N},\boldsymbol{\tau}^{cp}, \boldsymbol{\mu}^{cp}, \boldsymbol{\Sigma}^{cp}\}$. The problem then reads:
\begin{align} \label{obj_function_grouping_model}
\begin{split}
\max_{\mathcal{N}\in \mathcal{D}(N)} &\left[ \max_{\boldsymbol{\tau}^{cp}, \boldsymbol{\mu}^{cp}, \boldsymbol{\Sigma}^{cp}} \sum_{n=1}^N q_n \sum_{g=1}^G l_{ng} \log{\left(\tau_g^{cp} \phi(\mathbf{x}_n^{c}, x_n^p; \boldsymbol{\mu}_g^{cp}, \boldsymbol{\Sigma}_g^{cp} \right)} \right. +\\
&\quad+\left. \max_{\boldsymbol{\psi}_1,\ldots,\boldsymbol{\psi}_N} \sum_{n=1}^N(1-q_n)\log{w(\mathbf{x}_n^{c}, x_n^p , \mathbf{l}_n;\boldsymbol{\psi}_n)}\right].
\end{split}
\end{align}
By property \eqref{sep_condition}, any configuration that maximizes the first addend in \eqref{obj_function_grouping_model} also maximizes the second one. For a fixed partition $\mathcal{N}\in \mathcal{D}(N)$, the MLE for the first quantity are given by:
\[\hat{\tau}^{cp}_g=\frac{\sum_{n=1}^N q_nl_{ng}}{\lceil N(1-\gamma)\rceil}\:\:\:\:\: g=1,\ldots, G \]
\[\hat{\boldsymbol{\mu}}^{cp}_g=\frac{\sum_{n=1}^N q_nl_{ng}(\mathbf{x}_n^{c}, x_n^p)}{\sum_{n=1}^Nq_nl_{ng}}\:\:\:\:\: g=1,\ldots, G.\]
Estimation of $\boldsymbol{\Sigma}_g^{cp}$ depends on the considered patterned model, details are given in \cite{Bensmail1996}. Operatively, the final estimates are obtained via a REDDA model fitted on $\mathbf{x}_n^{c}, x_n^p$, see Section \ref{sec:REDDA_varsel}.
\subsection*{No grouping Model}
The log-likelihood function of the spurious outliers model under the no grouping structure is:
\begin{align} \label{ll_no_grouping_model_appendix}
\begin{split}
\ell(\mathcal{D},\boldsymbol{\tau}^{c}, \boldsymbol{\mu}^{c}, \boldsymbol{\Sigma}^{c}, \alpha, \boldsymbol{\beta}, \sigma^2)=&\sum_{n=1}^N q_n \sum_{g=1}^G l_{ng} \log{\left[\tau_g^{c} \phi(\mathbf{x}_n^{c}; \boldsymbol{\mu}_g^{c}, \boldsymbol{\Sigma}_g^{c})\right]}+\\
+&\sum_{n=1}^Nq_n\log{\left[ \phi(x^p_n; \alpha+\boldsymbol{\beta}^{'}\mathbf{x}^r_n, \sigma^2)\right]}+\\
+&\sum_{n=1}^N(1-q_n)\log{w(\mathbf{x}_n^{c}, x_n^p;\boldsymbol{\psi}_n)}
\end{split}
\end{align}
to be maximized with respect to $\{\mathcal{N},\boldsymbol{\tau}^{c}, \boldsymbol{\mu}^{c}, \boldsymbol{\Sigma}^{c}, \alpha, \boldsymbol{\beta}, \sigma^2\}$. The problem then reads:
\begin{align} \label{obj_function_no_grouping_model}
\begin{split}
\max_{\mathcal{N}\in \mathcal{D}(N)} &\left[ \max_{\boldsymbol{\tau}^{c}, \boldsymbol{\mu}^{c}, \boldsymbol{\Sigma}^{c}} \sum_{n=1}^N q_n \sum_{g=1}^G l_{ng} \log{\left[\tau_g^{c} \phi(\mathbf{x}_n^{c}; \boldsymbol{\mu}_g^{c}, \boldsymbol{\Sigma}_g^{c})\right]} \right.+\\
&\quad+ \max_{\alpha, \boldsymbol{\beta}, \sigma^2} \sum_{n=1}^Nq_n\log{\left[ \phi(x^p_n; \alpha+\boldsymbol{\beta}^{'}\mathbf{x}^r_n, \sigma^2)\right]} +\\
&\quad+\left. \max_{\boldsymbol{\psi}_1,\ldots,\boldsymbol{\psi}_N} \sum_{n=1}^N(1-q_n)\log{w(\mathbf{x}_n^{c}, x_n^p , \mathbf{l}_n;\boldsymbol{\psi}_n)}\right].
\end{split}
\end{align}
By property \eqref{sep_condition}, any configuration that maximizes the sum of the first and second term in \eqref{obj_function_no_grouping_model} also maximizes the third one. For a fixed partition $\mathcal{N}\in \mathcal{D}(N)$, the first two quantities can be separately maximized, leading to the following MLE
\[\hat{\tau}^{c}_g=\frac{\sum_{n=1}^N q_nl_{ng}}{\lceil N(1-\gamma)\rceil}\:\:\:\:\: g=1,\ldots, G \]
\[\hat{\boldsymbol{\mu}}^{c}_g=\frac{\sum_{n=1}^N q_nl_{ng}\mathbf{x}_n^{c}}{\sum_{n=1}^Nq_nl_{ng}}\:\:\:\:\: g=1,\ldots, G.\]
for the former term, where as usual $\hat{\boldsymbol{\Sigma}}_g^{c}$ depends on the considered patterned model. ML estimates for the regression coefficients are obtained solving the following minimization problem:
\begin{equation}
\min_{\alpha, \boldsymbol{\beta}} \sum_{n=1}^N q_n (x_n^p-\alpha-\boldsymbol{\beta}^{'}\mathbf{x}_n^r)^2
\end{equation}
which is very similar to the least trimmed squares method \citep{Rousseeuw1984}. Lastly, the variance is estimated as follows:
\[\hat{\sigma}^2=\frac{1}{\lceil N(1-\gamma)\rceil}\sum_{n=1}^Nq_n(x_n^p-\hat{\alpha}-\hat{\boldsymbol{\beta}}^{'}\mathbf{x}_n^r).\]
Operatively, the MLE for \eqref{ll_no_grouping_model_appendix} are obtained combining a REDDA model on $\mathbf{x}_n^{c}$ with a robust linear regression of $x^p_n$ on $\mathbf{x}_n^{r}$. The discriminating function in \eqref{D_NOgroup_2} is used to determine the subset of untrimmed units on which to compute the estimates defined above, iterating the algorithm until the same observations are discarded in two consecutive steps. Lastly, at each iteration, similarly to what performed in \texttt{clustvarsel} \citep{Scrucca2018}, the subset of variables $\mathbf{x}_n^{r}$ is determined with the \texttt{bicreg} function in the \texttt{BMA R} package \citep{Raftery2018}.

\section{Further aspects for the ML subset selector approach} \label{sec:appendix_B} 
This final Section discusses the computational details of the algorithm used for fitting the ML subset selector, whose main steps are reported in Section \ref{sec:EMST}. For achieving flexibility, parsimony and computational speed, the family of patterned models based on the eigenvalue decomposition in \eqref{sigma_dec_varsel} of \cite{Bensmail1996} is considered. Particularly, we adopt the three-letter identifier used in the \texttt{mclust} software for naming the models, where the volume, shape and orientation can be either equal (E) or different (V) across groups, with full (**E, **V), diagonal (**I) or spherical (*II) components: we refer to \cite{Scrucca2016} for the complete details. Let us further introduce the following notations: for a $d \times d$ matrix $\boldsymbol{A}$, $\mbox{diag}(\boldsymbol{A})$ denotes the $d \times d$ diagonal matrix whose diagonal entries are the same of the matrix $\boldsymbol{A}$. Lastly, $\boldsymbol{A}(i,j)$ denotes the scalar entry at the $i$th row and $j$th column of the matrix $\boldsymbol{A}$. 

\subsection*{Computational details on the M-step}
As previously mentioned, we refer the reader to \cite{Bensmail1996} for a complete treatment on the estimation of $\bSigma_g$, $g=1,\ldots,G$ under the 14 covariance structures. Conditioning on the chosen model, the estimation of the pooled covariance matrix $\bSigma$ has the following form:
\begin{itemize}
\item Ellipsoidal: \[\hat{\bSigma}_{ell}=\frac{1}{\lceil N(1-\gamma)\rceil}\sum_{n=1}^N\zeta(\mathbf{x}_n)\left[(\mathbf{x}_n-\hat{\bmu})(\mathbf{x}_n-\hat{\bmu})^{'}\right]\]
for EEE, VEE, EVE, EEV, VVE, VEV, EVV and VVV models
\item Diagonal: \[\hat{\bSigma}_{diag}=\mbox{diag}(\hat{\bSigma}_{ell})\]
for EEI, VEI, EVI, VVI models. 
\item Spherical: \[\hat{\bSigma}= \frac{1}{P}\sum_{d=1}^P \hat{\bSigma}_{diag}(d,d)\boldsymbol{I}_P\]
for EII, VII models
\end{itemize}
\subsection*{Computational details on the S-step}
The S-step involves a discrete structure optimization, where we seek to determine the set of $p$ variables that minimizes \eqref{s-step_varsel}. Solving the problem by exhaustive enumeration is feasible only when ${P \choose p}$ is not too large, sadly it is rarely the case in a high-dimensional setting. Thus, the considered implementation relies on a stochastic algorithm for fixed-size subset selection, by means of the \texttt{kofnGA R} package \citep{Wolters2015}. Nonetheless, for specific patterned structures, simpler form of the objective function may be derived: see the following sections.
\subsubsection*{EEE model}
For the homoscedastic model (EEE), \eqref{s-step_varsel} simplifies as follows:
\begin{equation}
h(F)=\log \det \hat{\bSigma}_{EEE,F}-\log \det \hat{\bSigma}_{ell,F}
\end{equation}
where
\[\hat{\bSigma}_{EEE,F}=\frac{1}{\lceil N(1-\gamma)\rceil}\sum_{g=1}^G \hat{n}_g \sum_{n=1}^N\zeta(\mathbf{x}_{n})\left[(\mathbf{x}_{n,F}-\hat{\bmu}_{g,F})(\mathbf{x}_{n,F}-\hat{\bmu}_{g,F})^{'}\right]\]
and
\[\hat{n}_g=\frac{\sum_{n=1}^N \zeta(\mathbf{x}_n)l_{ng}}{\lceil N(1-\gamma)\rceil}.\]
It is nevertheless computationally efficient to derive $\hat{\bSigma}_{EEE}$ for the full dimension $P$ at once and to extract the sub-matrix $\hat{\bSigma}_{EEE,F}$ when needed.
\subsubsection*{VVI model}
For the heteroscedastic diagonal model (VVI), \eqref{s-step_varsel} simplifies to:
\begin{equation}
h(F)=\sum_{k \in F}\sum_{g=1}^G\hat{\tau}_g \log \frac{\hat{\bSigma}_{g}(k,k)}{\hat{\bSigma}_{diag}(k,k)}
\end{equation}
for which $\hat{F}$ is the set of the indices $k$ with the $p$ smallest sums $\sum_{g=1}^G\hat{\tau}_g \log \frac{\hat{\bSigma}_{g}(k,k)}{\hat{\bSigma}_{diag}(k,k)}$.
\subsubsection*{EEI model}
For the homoscedastic diagonal model (EEI), \eqref{s-step_varsel} reads:
\begin{equation}
h(F)=\sum_{k \in F}\log \frac{\hat{\bSigma}_{EEI}(k,k)}{\hat{\bSigma}_{diag}(k,k)}
\end{equation}
with
\[\hat{\bSigma}_{EEI}=\frac{1}{\lceil N(1-\gamma)\rceil}\sum_{g=1}^G \hat{n}_g \mbox{diag} \left(\sum_{n=1}^N\zeta(\mathbf{x}_{n})\left[(\mathbf{x}_{n,F}-\hat{\bmu}_{g,F})(\mathbf{x}_{n,F}-\hat{\bmu}_{g,F})^{'}\right]\right).\]
In this case, $\hat{F}$ is the set of the indices $k$ with $p$ smallest quotients $\frac{\hat{\bSigma}_{EEI}(k,k)}{\hat{\bSigma}_{diag}(k,k)}$.

\subsection*{Computational details on the T-step}
When the full dimension $P$ is large, it may occur that $\hat{\bSigma}_{\hat{E}|\hat{F}}$ is not of full rank. In this case, it is still possible to estimate a singular normal distribution on a subspace of the set $\hat{E}$ of irrelevant variables. The associated density will then be:
\begin{equation}
\frac{(2 \pi)^{-k / 2}}{\left(\prod_{k=1}^K\omega_{k}\right)^{1 / 2}} \exp \left\{-\frac{1}{2}(\mathbf{x}_{n,\hat{E}}-\hat{\boldsymbol{G}}_{\hat{E}|\hat{F}}\mathbf{x}_{n,\hat{F}}-\hat{\bmu}_{\hat{E}|\hat{F}})^{\prime} \hat{\bSigma}_{\hat{E}|\hat{F}}^{-}(\mathbf{x}_{n,\hat{E}}-\hat{\boldsymbol{G}}_{\hat{E}|\hat{F}}\mathbf{x}_{n,\hat{F}}-\hat{\bmu}_{\hat{E}|\hat{F}})\right\}
\end{equation}
where $\hat{\bSigma}_{\hat{E}|\hat{F}}^{-}$ is the g-inverse of $\hat{\bSigma}_{\hat{E}|\hat{F}}$ and $\omega_{1}, \ldots, \omega_{K}$ are the non-zero eigenvalues of $\hat{\bSigma}_{\hat{E}|\hat{F}}$.
\subsection*{Models comparison}
As a final remark, we mention the possibility of developing a procedure for automatically choosing the best model within the 14 parsimonious structures in the ML subset selector approach. One could rely on a BIC-like criterion \citep{Schwarz1978}, penalizing twice the final maximized trimmed log-likelihood by the number of estimated parameters and untrimmed observations, retaining the model that presents the highest value. However, this would rapidly increase the computational time needed for performing the analysis. For this reason, in both the simulation study and in the application, a VVV model only was considered when fitting the ML subset selector approach.

\end{document}